\documentclass[aps,prb,10pt,twocolumn,floatfix,showpacs,superscriptaddress,longbibliography]{revtex4-1}
\usepackage[colorlinks=true,citecolor=blue,linkcolor=blue,urlcolor=blue]{hyperref}
\usepackage[normalem]{ulem}
\usepackage{amsmath}
\usepackage{amssymb}
\usepackage{graphicx,amsmath,amsfonts,bm}
\usepackage{epstopdf}
\usepackage{bm}
\usepackage{bbm}
\usepackage{color}
\usepackage{relsize}
\usepackage{dsfont}
\usepackage{braket}
\usepackage[caption=false]{subfig}

\usepackage[T1]{fontenc}
\usepackage{dsfont}
\usepackage{mathbbol}
\usepackage{hyperref}
\renewcommand{\i}{\ensuremath{\mathrm{i}}}

\newcommand{\diag}{\mathrm{diag}}
\newcommand{\sgn}{\operatorname{{\mathrm{sgn}}}}

\newcommand{\de}{\ensuremath{\mathrm{d}}\!}
\DeclareMathOperator{\Pf}{Pf}
\allowdisplaybreaks

\begin{document}
\title{Confinement-induced Majorana modes in a nodal topological superconductor}

\author{Simone \surname{Traverso}} \email{simone.traverso@edu.unige.it}
\affiliation{Dipartimento di Fisica, Università degli studi di Genova, Via Dodecaneso 33, Genova, 16146, Italia}
\affiliation{SPIN-CNR, Consiglio Nazionale delle Ricerca,Via Dodecaneso 33, Genova, 16146, Italia}
\author{Niccolò \surname{Traverso Ziani}} \email{niccolo.traverso.ziani@unige.it}
\affiliation{Dipartimento di Fisica, Università degli studi di Genova, Via Dodecaneso 33, Genova, 16146, Italia}
\affiliation{SPIN-CNR, Consiglio Nazionale delle Ricerca,Via Dodecaneso 33, Genova, 16146, Italia}
\author{Maura \surname{Sassetti}} \email{maura.sassetti@unige.it}
\affiliation{Dipartimento di Fisica, Università degli studi di Genova, Via Dodecaneso 33, Genova, 16146, Italia}
\affiliation{SPIN-CNR, Consiglio Nazionale delle Ricerca,Via Dodecaneso 33, Genova, 16146, Italia}
\author{Fernando \surname{Dominguez}}\email{f.dominguez@tu-braunschweig.de}
\affiliation{Faculty of Physics and Astrophysics and W\"urzburg-Dresden Cluster of Excellence ct.qmat, University of W\"urzburg, 97074 W\"urzburg, Germany}
\affiliation{Institut f\"{u}r Mathematische Physik, Technische Universit\"{a}t Braunschweig, Mendelssohnstraße 3, Braunschweig, 38106, Germany}

\date{\today}

\begin{abstract}
We investigate the topological phase diagram of {an extension of the Haldane model with equal spin pairing superconductivity}. In two dimensions, we find a topological nodal superconducting phase, which exhibits a chiral Majorana mode propagating along the edges of nanoribbons with cylindrical boundary conditions. This phase is however unstable in a finite two-dimensional rectangular-shaped lattice, yielding corner states close to zero energy in a flake with alternating zigzag and armchair edges. 
When we reduce one of the dimensions, quantum confinement gaps out the bulk bands faster than the edge states. In this scenario, hybridization between the edge states can then result in Majorana zero modes.
Our results hence suggest quantum confinement as a crucial ingredient in building quasi-one-dimensional topological superconducting phases out of two-dimensional nodal topological superconductors.
Furthermore, we characterize the emergence of this novel topological phase by means of its topological invariant, coinciding with a quantized conductance of $2 e^2/h$ in a normal-superconducting junction.
\end{abstract}

\maketitle

\section{Introduction}
\label{sec:intro}
Majorana bound states (MBS) emerge as zero energy excitations at the boundaries of one-dimensional topological superconductors. These quasiparticles fulfill the usual fermionic anticommutation relations $\{\gamma, \gamma^\dagger\}=1$~\cite{Read2000a, Ivanov2001a,Kitaev_2001}. However, in contrast to regular fermions, their creation and annihilation operators are equal, i.e.~$\gamma=\gamma^\dagger$. This unusual property leads to a neutral charge, a badly-defined occupation number and gives rise to anomalous signatures present, for example, in the electrical \cite{Law2009a} and thermal \cite{Wimmer2010a, Akhmerov2011a} conductance, shot-noise \cite{Akhmerov2011a}, fractional Josephson effect \cite{Kitaev_2001}, etc. 
Apart from their fundamental interest, they have potential applications, particularly in the protection of quantum information and fault-tolerant quantum computation \cite{Ivanov2001a}, see further information in Refs.~\cite{Nayak2008a, Alicea2012a, Beenakker2013a, Aguado2017a, Schuray2020a, Prada2020a, Flensberg2021a}.

Among all possible candidates, $p$-wave superconductors are the most widespread platforms hosting topological superconductivity~\cite{Kitaev_2001}. However, the lack of this superconducting phase in nature has forced researchers to look for alternative routes. 
Early proposals to overcome this difficulty suggested a conventional proximity effect on one-dimensional spin-momentum locking states directly on a quantum spin Hall edge~\cite{Fu2009a}, or on Rashba quantum wire under Zeeman fields~\cite{Lutchyn2010a,Oreg2010a}. Ever since, a huge experimental effort has been made to bring these systems to the laboratory, where signatures of zero energy modes compatible with the presence of MBS were found, for example, in the conductance~\cite{Mourik2012a, Deng2012a, Das2012a, Nichele2017a} and the fractional Josephson effect~\cite{Rokhinson2012a, Wiedenmann2016a, Li2018a, Bocquillon2016a, Deacon2017a, Laroche2019a}. Unfortunately, the identification of these zero energy modes with Majorana bound states is not univocal as the presence of quasi-Majorana bound states~\cite{Liu2017a, Fleckenstein2018a, Moore2018a, Marra2019a, Dmytruk2020a} or trivial zero-energy Andreev bound states~\cite{Cayao2017a,Oladunjoye2019a} can also give rise to similar features. 
Hence, the increasing interest for detection schemes that allow to distinguish among both scenarios~\cite{Haim2015a, Fleckenstein2021a, Pakizer2021a, Dominguez2024a, Bittermann2024a} and alternative platforms for topological superconductivity that are less sensible to disorder, like for example, atomic chains~\cite{Perge2014a, Jeon2017a}, two-dimensional semiconductor heterostructures~\cite{Suominen2017a, Nichele2017a}, quantum Hall edge states~\cite{San-Jose2015a, Finocchiaro2018a}, bilayer graphene~\cite{Penaranda2023a}, {honeycomb lattice based nanostructures}~\cite{Ribeiro_2022,Ribeiro_2023}.

In this contribution, we investigate a new mechanism leading to the emergence of Majorana bound states, {based on the geometrical confinement of a nodal topological superconductor. To illustrate this mechanism, we consider an extension of the Haldane model for a Chern insulator with equal spin pairing (ESP) superconductivity. Here, we show that a $\mathbb{Z}_2$ topological nodal superconducting phase arises from the interplay between the trivial masses of the model with the ESP superconducting coupling. In this phase,} the model exhibits a chiral Majorana mode in zigzag or armchair nanoribbons with cylindrical boundary conditions. 
Interestingly, corner states close to zero energy arise when considering a finite rectangular-shaped lattice. These corner states result from the alternating zigzag and armchair boundary conditions, which impose an infinite (finite) localization length on the chiral mode along the zigzag (armchair) edges. Similarly to higher-order topological systems, a topological armchair edge becomes surrounded by two effective trivial edges, and thus, the initial chiral Majorana mode becomes localized at the corners of the flake. In this scenario, Majorana bound states can emerge at the ends of a quasi-one dimensional zigzag nanoribbon, resulting from the reduction of the armchair directions. Here, the mass confinement plays a crucial role as it opens a gap in the bulk spectrum, providing a finite localization length on the chiral Majorana modes hosted by the zigzag edges. In narrow zigzag nanoribbons, the re-established edge states can thus hybridize and give rise to a complex, width-dependent phase diagram, whose topological regions host quasi-zero dimensional MBSs. We characterize the topological phase by means of the calculation of the topological invariant and the conductance quantization, finding an exact correspondence. Finally, we discuss possible connections with experimental platforms.

The paper is organized as follows. First, in Sec.~\ref{sec:model} we introduce the Haldane model together with the spinless superconducting model used in the rest of the paper. Furthermore, we provide insight on the bulk spectrum and the parameter regime at which the superconducting spectrum develops nodal points. Then, in Sec.~\ref{sec:topo2d}, we introduce a $\mathbb{Z}_2$ topological invariant to characterize a topological nodal superconducting phase emerging in the presence of particle-hole and inversion symmetry. Furthermore, we show the one dimensional edge states with zero-energy Majorana solutions that emerge when bringing the bulk Hamiltonian into a lattice with armchair and/or zigzag boundary conditions. In Sec.~\ref{Sec:Q1D}, we complete our analysis studying the topological phase diagram of a quasi-one dimensional zigzag nanoribbon. To this aim we introduce a topological invariant, the Majorana number, adapted from the one-dimensional Kitaev Hamiltonian. We compare these results with spectrum and zero temperature conductance calculations, finding an exact correspondence between them. Finally, we conclude in Sec.~\ref{Sec:conclusions} discussing our results.

\section{Model and bulk phase diagram}
\label{sec:model}

\subsection{Bulk Hamiltonian}

\begin{figure}
\centering
\includegraphics[width=\linewidth]{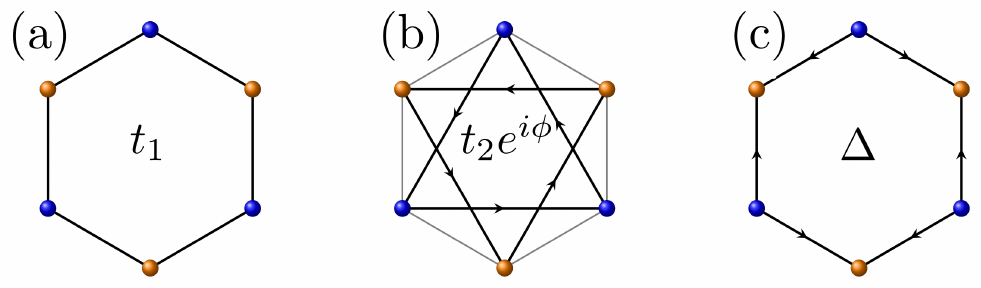}
    \caption{(a) First neighbor real hopping, with amplitude $t_1$. (b) Second neighbor complex hopping, with amplitude $t_2$ and a phase $e^{i\phi}$ ($e^{-i\phi}$)  associated to the hoppings in the (opposite) direction of the arrows. (c) First neighbor superconducting hoppings with amplitude $\Delta$: a hole is destroyed at the tail of each arrow, and an electron is created at the corresponding end.}
    \label{fig:hopping_scheme}
\end{figure}

The model considered in this contribution, consists on the combination of two models responsible for the emergence of two representative classes of topological states: the Haldane model~\cite{Haldane_1988} and the {ESP (spinless) superconductor}~\cite{Kitaev_2001}. To this aim, we set first the honeycomb lattice with atoms of sublattice $A$ and $B$ placed at positions
\begin{equation}
 \mathbf{R}_{ln}^A = l\mathbf{a}_{1} + n\mathbf{a}_{2}, \qquad \mathbf{R}_{ln}^B = l\mathbf{a}_{1} + n\mathbf{a}_{2} + a\left(\frac{1}{2},\frac{1}{2\sqrt{3}}\right),
\end{equation}
with the primitive vectors chosen as $\mathbf{a}_{1} = a(1,0)$, $\mathbf{a}_{2} = a\left(1/2, \sqrt{3}/2\right)$, $a$ the second neighbor distance.

{On this lattice, we consider the real space tight-binding Hamiltonian}
\begin{equation}
	\begin{split}
		H &= \sum_{l,n}\Big[ t_1  (a^\dagger_{ln} b_{ln} + a^\dagger_{ln} b_{l-1n}+a^\dagger_{ln}b_{ln-1})+\text{h.c.}\\    
		&+t_2 e^{+i\phi} (a_{ln}^\dagger a_{l+1n} + a^\dagger_{ln}a_{l-1n+1}+a^\dagger_{ln}a_{ln-1})+\text{h.c.}\\
		&+t_2 e^{-i\phi}( b_{ln}^\dagger b_{l+1n} + b^\dagger_{ln}b_{l-1n+1}+b^\dagger_{ln}b_{ln-1})+\text{h.c.}\\
		&+\Delta (a^\dagger_{ln} b^\dagger_{ln} + a^\dagger_{ln} b^\dagger_{l-1n}+ a^\dagger_{ln}b^\dagger_{ln-1})+\text{h.c.}\\
		&+ (\mu+m)a^\dagger_{ln}a_{ln}+(\mu-m)b^\dagger_{ln}b_{ln}\Big],
	\end{split}
    \label{eq:real_space_ham}
\end{equation}
{with the operators $a_{ln}$ ($b_{ln}$) destroying a fermion at the position $\mathbf{R}_{ln}^A$ ($\mathbf{R}_{ln}^B$) and satisfying canonical anti-commutation relations.} The model exhibits three types of hopping amplitudes, illustrated in real space in Fig.~\ref{fig:hopping_scheme}: (a) graphene nearest neighbor (NN) hopping, with amplitude $t_1$; (b) next-nearest neighbor (NNN) hopping, with amplitude $t_2$, and a phase $\phi$ resulting from a {staggered magnetic field with the same symmetries of the lattice, and zero net flux} over the lattice unit cell~\cite{Haldane_1988}; 
(c) nearest neighbor equal spin pairing superconducting hopping, with amplitude $\Delta$. {Furthermore, it presents both a staggered mass term, assuming opposite values ($\pm m$) on the sublattices $A$ and $B$, and a chemical potential ($\mu$) term.} In what follows, we fix $\phi =\pi/2$, set $a$ as the unit of length, {and $t_1$ as the (unitary) energy scale. Moreover, we} assume the superconducting gap parameter $\Delta$ to be real. Furthermore, since our study is conceptual, we fix $t_2=0.3 t_1$ and study the resulting phase diagrams for a general set of parameters $(m,\Delta)$, with $\mu=0$ or $(\mu,\Delta)$, with $m=0$.

{Regarding the choice of the mean-field superconducting term, models presenting ESP superconductivity on bipartite lattices have already been considered in literature}~\cite{Chen_2018, Ezawa_2018, Zhang_2019}. {In particular, it has been argued that spin-triplet $f$-wave superconductivity could be induced in honeycomb materials}~\cite{Ezawa_2018}. {Indeed, it is fully compatible with the lattice symmetries, which is not the case, for example, for purely $p$-wave superconductivity. On the honeycomb lattice, ESP $f$-wave superconducting pairing would manifest in both a NN and a NNN superconducting hopping term}~\cite{Ezawa_2018, Zhang_2019}. {Here however, we neglect the latter. The motivation behind this choice is two-fold: first, in any real-world setup the typical strength of the proximity induced superconductivity is already much weaker than the other relevant coupling terms in the Hamiltonian, and an NNN coupling is expected to be even weaker than a first neighbour one. Second, we do not expect the addition of an NNN term to qualitatively alter our results.
Unlike previous works on similar systems where exact solution along critical lines in presence of Hubbard interactions were sought}~\cite{Chen_2018, Ezawa_2018, Zhang_2019}, {in this contribution we focus on exploring the emergence of a topological superconducting regime in a proximitized Chern insulator. Moreover, we show that this regime results from the interplay between ESP superconductivity and trivial masses.}

\begin{widetext}

\begin{figure}[h]
    \centering
    \includegraphics[width=\linewidth]{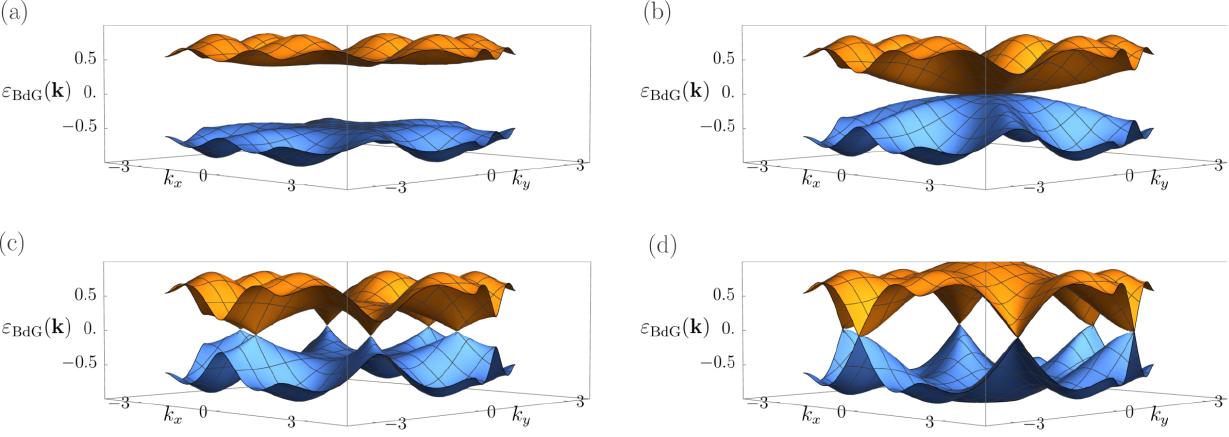}
    \caption{Low energy BdG band spectrum for the cases $\mu=0, \ m=1$ at: (a) $\Delta=0.9<\Delta_{\text{c},1}$; (b) $\Delta=\sqrt{10}/3=\Delta_{\text{c},1}$ (c) $\Delta_{\text{c},1} < \Delta = 1.2 < \Delta_{\text{c},2}$; (d) $\Delta = \sqrt{2} = \Delta_{\text{c},2}$. The other parameters are set as in the main text, \emph{i.e.}~$t_1=1,\ t_2=0.3$.}
    \label{fig:bulk-bands}
\end{figure}

In the bulk, the Bogoliubov de Gennes (BdG) Hamiltonian {of the NN-ESP} superconducting Haldane model is given by
$H = \frac 12 \sum_{\mathbf k} \Psi^\dagger(\mathbf k) \mathcal H(\mathbf k) \Psi(\mathbf k)$,
with
\begin{equation}
    \mathcal H(\mathbf k)=
    \begin{bmatrix}
        m+\mu+2t_2 S(\mathbf k)  & t_1 f(\mathbf k) &0 &\Delta f(\mathbf k) \\
        t_1 f^{\ast}(\mathbf k)& -m+\mu-2t_2 S(\mathbf k)&-\Delta f^\ast(\mathbf k) & 0 \\
        0& -\Delta f(\mathbf k) &-m-\mu+2t_2 S(\mathbf k) & -t_1 f(\mathbf k) \\
        \Delta f^\ast(\mathbf k) & 0 &-t_1 f^{\ast}(\mathbf k) & m-\mu-2t_2 S(\mathbf k)
    \end{bmatrix}.
    \label{eq:Hbdg}
\end{equation}
Here $f(\mathbf k)= ( 1 + e^{-i\mathbf k \cdot \mathbf a_1}+e^{-i\mathbf k \cdot \mathbf a_2})$ and $S(\mathbf k) = (-\sin(\mathbf k \cdot \mathbf a_1) + \sin(\mathbf k \cdot( \mathbf a_1-\mathbf a_2))+\sin(\mathbf k \cdot \mathbf a_2))$ arise from the nearest and next-nearest neighbor hoppings respectively. The spinor 
$\Psi(\mathbf k)=\left[ a(\mathbf k),b(\mathbf k),a^\dagger(-\mathbf k),b^\dagger(-\mathbf k)\right]^T$, with {the fermionic} annihilation operators $a(\mathbf k)$ and $b(\mathbf k)$ acting on sublattice $A$ and $B$ with momentum $\mathbf k$. A complete derivation of the bulk BdG Hamiltonian, starting from the real space one given in Eq.~\eqref{eq:real_space_ham}, can be found in App.~\ref{app.SN1}. 

\subsection{Bulk spectrum}

Henceforth, we restrict our analysis to the situation where either $\mu$ or $m$ are finite, i.e.~for $\mu=0$ and $m\neq0$ or $\mu\neq0$ and $m=0$. In these cases, the spectrum exhibits an additional inversion symmetry, see App.~\ref{app.SN1}, which, together with particle-hole symmetry, yields a symmetric spectrum {with} respect to zero energy. 
Under these conditions, we can solve for the bulk spectrum by squaring the Hamiltonian twice. Specifically, one finds the following expression for the (squared) bulk bands for $\mu=0$
\begin{equation}
    \varepsilon^2(\mu=0;\mathbf k) = m^2+4t_2^2S^2(\mathbf k) +(t_1^2+ \Delta^2) |f(\mathbf k)|^2\pm
    \\
    \sqrt{[4mt_2S(\mathbf k)]^2 +[2m\Delta]^2|f(\mathbf k)|^2+[2t_1\Delta|f(\mathbf k)|^2]^2},
    \label{eq:bands_mu_0}
\end{equation}
and for $m=0$
\begin{equation}
    \varepsilon^2(m=0;\mathbf k) = \mu^2+4t_2^2S^2(\mathbf k) +(t_1^2+ \Delta^2) |f(\mathbf k)|^2\pm\sqrt{[4\mu t_2S(\mathbf k)]^2 +[2\mu t_1]^2|f(\mathbf k)|^2+[2t_1\Delta|f(\mathbf k)|^2]^2}.
    \label{eq:bands_m_0}
\end{equation}
Crucially, the spectra in the two cases $\mu=0$ and $m=0$ coincide upon exchanging of $\mu \leftrightarrow m$ and $t_1\leftrightarrow \Delta$. This corresponds to the canonical transformation $b_{ij}\leftrightarrow b^\dagger_{ij}$ for $i,j$ in the model Hamiltonian. 

\end{widetext}

In Fig.~\ref{fig:bulk-bands}, we represent the low energy BdG bands of Eq.~\eqref{eq:Hbdg} for $\mu=0$ and $m=1$ and different values of $\Delta$, exhibiting a gapped spectrum (a) or a nodal superconducting phase with one (b) or six nodal points (c) and (d). We can predict the appearance of these nodal points in terms of the model parameters $\Delta, t_1, t_2, \mu$ and $m$ and then map our conclusions to the $m=0$ case by exchanging $\mu \leftrightarrow m$ and $t_1\leftrightarrow \Delta$. 

Starting from the analytic expression of the energy bands reported in Eq.~(\ref{eq:bands_mu_0}), one can identify the gapped and gapless regions of the bulk phase diagram. First, one notices that at the Dirac points ($\mathbf{k}_\text{D} = \pm \frac{4\pi}{3}(1,0)$), $f(\mathbf k)=0$. Thus, we retrieve the Haldane gap closing condition,
\begin{equation}
	m = \pm 2t_2S(\mathbf{k}_{\text{D}}) = \pm  3\sqrt{3} t_2.
\end{equation}
In this sense, the equal spin superconducting pairing considered here, being proportional to $f(\mathbf k)$ in momentum space, does not compete with the Haldane mass at the Dirac point. In other words, a superconducting gap cannot be opened at the Dirac points in the present model.

Furthermore, the spectrum presents a nodal superconducting phase for the superconducting amplitudes between the critical lines
\begin{align}
    \Delta_{\text{c},1} &=\sqrt{t_1^2+(\Theta/3)^2} \text{ ~~and}\\ 
    \Delta_{\text{c},2} &=\sqrt{t_1^2+\Theta^2},
\end{align}
which are plotted in red in Fig.~\ref{fig:Bulk-PD}(a,b), with  $\Theta=m$ or $i \mu$.
For $\Delta=\Delta_{\text{c},1}$, the gap closes at the $\Gamma$-point [Fig.~\ref{fig:bulk-bands}(b)], and develops six nodal points that shift towards the $M$-points as we tune $\Delta$ from $\Delta_{\text{c},1}$ to $\Delta_{\text{c},2}$ [Fig.~\ref{fig:bulk-bands}(c),~(d)]. Then, for $\Delta=\Delta_{\text{c},2}$ the gap reopens at the $M$ points. This behavior is diagrammatically represented in Fig.~\ref{fig:Bulk-PD}(c). The equations of the critical lines $\Delta_{\text{c},1}$ and $\Delta_{\text{c},2}$, together with an implicit expression of the nodal points as a function of the model parameters are derived in App.~\ref{app.nodal}. 

\section{Topological phase diagram}
\label{sec:topo2d}

\begin{figure}    
\centering
\includegraphics[width=\linewidth]{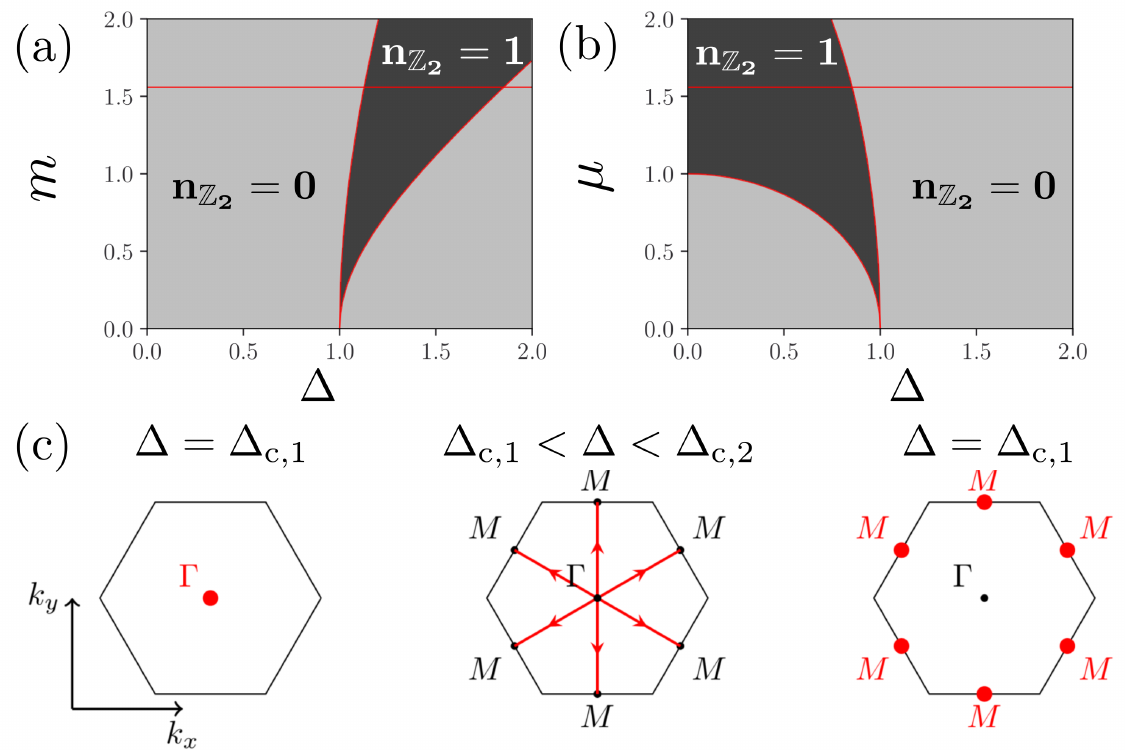}
\caption{(a,b) Topological bulk phase diagram for the case $\mu=0$ and $m=0$ respectively. The darker regions correspond to a gapless bulk, with $n_{\mathbb{Z}_2}=1$, representing a nodal topological superconducting phase. The lighter regions instead correspond to a gapped bulk with $n_{\mathbb{Z}_2}=0$. The red lines mark the boundaries of the regions where the gap is expected to be zero. {We set $t_2=0.3$ in both panels.} (c) Scheme of the Brillouin zone with the points where the bulk dispersion relation is gapless highlighted for the different values of $\Delta$.}
\label{fig:Bulk-PD}
\end{figure}

We start with the topological characterization of the {NN-ESP} superconducting Haldane model described by the Bogoliubov de Gennes bulk Hamiltonian given by Eq.~\eqref{eq:Hbdg}.
The Haldane model without superconductivity, exhibits a Chern insulating phase with Chern number $\mathcal{C}=1$ for $|m|<3\sqrt{3}t_2$ and a trivial insulating phase for $|m|>3\sqrt{3}t_2$. The transition between both phases occurs at $m=3\sqrt{3}t_2$, where the gap closes at the $K/K'$-points.

In the presence of superconductivity, 
the Hamiltonian is in symmetry class D and admits a $\mathbb{Z}$ topological invariant, the Chern number $\mathcal{C}$,  which characterizes the topology of the gaped regimes. In this case, $\mathcal{C}=2$, which is inherited from the topology of the Haldane Hamiltonian. These gapped regions extend over $|\Theta|<3\sqrt{3}t_2$, with  $\Theta=m$ or $i \mu$, see the horizontal red lines in Fig.~\ref{fig:Bulk-PD}(a,b).
Thus, under open boundary conditions two pairs of chiral Majorana modes are found on each edge of the open system~\cite{Zhang_2019}. On the other hand, $\mathcal{C}=0$ in the gapped regions of the phase diagram satisfying $|\Theta|>3\sqrt{3}t_2$ and, correspondingly, under open boundary conditions no edge states occur in this phase.

The nodal superconducting phase also exhibits topologically protected edge states. In contrast to the gapped regions, the topological invariant cannot be defined globally, but in terms of momentum dependent topological numbers. To determine it, we need to take into account the position of the nodal points in the Brillouin zone and the codimension of the Hamiltonian, namely
\begin{equation}
    p=d_{\text{BZ}}-d_{n}=2,
\end{equation}
with the dimension of the Hamiltonian ($d_\text{BZ}=2$) and the dimension of the nodal point ($d_n=0$). Thus, $p$ together with the fact that the nodal points are placed away from high symmetry points, sets the topology of the Hamiltonian to be characterized by a $\mathbb{Z}_2$ invariant~\cite{Schnyder2015a}.

The topological invariant $\mathbb{Z}_2$ can be in general computed using the Wilson loop. Here, the idea is to choose a set of gapped Hamiltonians with $k$-values encircling  a nodal point, see App.~\ref{App.Wilson}. 
In this sense, the topological invariant signals whether the nodal point can shrink into points (trivial) or not (topological)~\cite{Sato2006a}.
We can simplify the calculation of the invariant by noticing that the nodal points lie along the lines connecting the $\Gamma$ and $M_i$ points, where we can define a path that is related by PHS, i.e.~$k\rightarrow -k$. Similarly as in the Kitaev model~\cite{Kitaev_2001, deJuan2014, Dominguez2022a}, for each $\Gamma-M$ line we can define a one-dimensional class-D $\mathbb{Z}_2$ invariant as the product of Pfaffians at the fixed points $\Gamma$ and $M_i$ points, namely
\begin{align}\label{eq.pff}
     (-1)^{n_{\mathbb{Z}_2}}=\text{sgn}\{ \text{Pf}[A({\boldsymbol \Gamma})] \text{Pf}[A( {\boldsymbol M_i})]\},
\end{align}
with the antisymmetric matrices $A=U H$ and $U$ the unitary operator part of the PHS operator $\mathcal{C}=U\mathcal{K}$.
Eq.~\eqref{eq.pff}, 
$\text{Pf}[A({\boldsymbol \Gamma})]=9 \Delta^2- 9 t_1^2-\Theta^2$, $\text{Pf}[A({\boldsymbol M}_i)]=\Delta^2-\Theta^2 - t_1^2 $ and $n_{\mathbb{Z}_2}=1(0)$ sets a non-trivial phase coinciding entirely with the nodal region delimited by $\Delta_{\text{c},2}\geq \Delta\geq \Delta_{\text{c},1}$, see Fig.~\ref{fig:Bulk-PD}(a).

\begin{widetext}

\begin{figure}
\centering
\includegraphics[width=\textwidth]{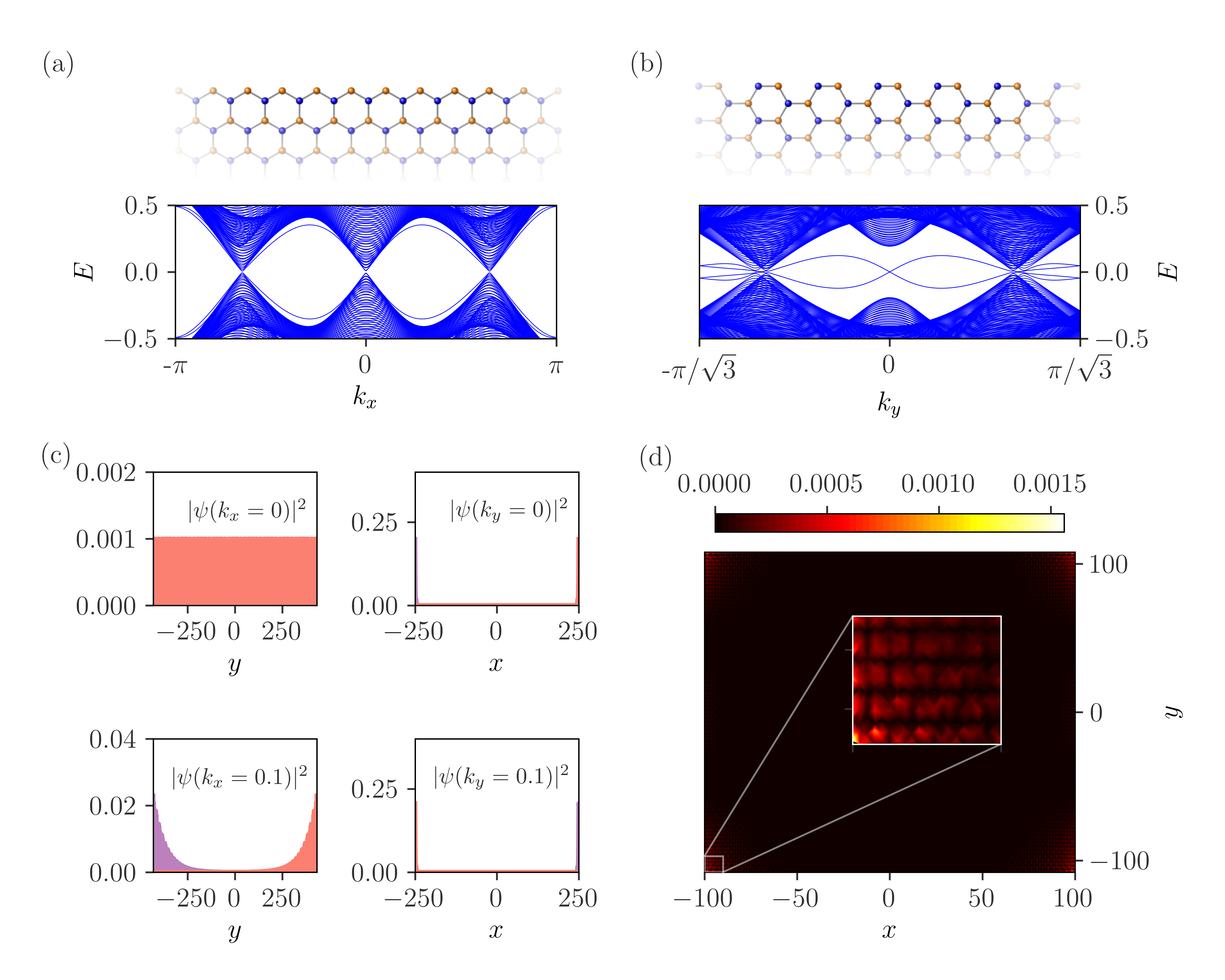}
\caption{(a) Bands of a wide zigzag nanoribbon (edge structure above) with $N_y=4\times 250$. (b) Bands of a wide armchair nanoribbon (edge structure above), with $N_x=2\times 250+1$. (c) 1D probability density profile associated to the lowest energy eigenstates of a zigzag nanoribbon with $k_x=0$ and $k_x=0.1/a$ in the left column and analogous comparison for an armchair nanoribbon in the right column. (d) Probability density plot associated to the lowest energy eigenstate of a rectangular-shaped flake, with an inset zooming on the bottom left corner, with $N_x=2\times 100+1$ and $N_y=4\times 55$. {In all  plots, the parameters are set as:} $\mu=0,\ m=1.75,\ \Delta = 1.4, \ t_2=0.3$.}
     \label{fig:Bulk_CMs}
\end{figure}

\end{widetext}

We confirm the presence of a topological nodal phase $(n_{\mathbb{Z}_2}=1)$ by calculating the energy spectrum of zigzag and armchair nanoribbons, see for example Fig.~\ref{fig:Bulk_CMs}(a,b) for $\mu=0$, $m=1.75$ and $\Delta=1.4$. Here, one can clearly see isolated energy modes connecting the projections of the bulk nodal points in the 1-dimensional Brillouin zone.

Surprisingly, when we set the Hamiltonian onto a two-dimensional lattice cropped with a rectangular shape, resulting in alternating zigzag and armchair boundaries, the local density of states at zero energy is not uniformly extended along the edges. In turn, it reveals the presence of corner states, see Fig.~\ref{fig:Bulk_CMs}(d). Here, the resemblance with higher order topological states origins from the fact that a chiral Majorana mode propagating along the armchair edges is surrounded by effectively trivial edge states on zigzag edges. Now, to answer why the zigzag modes become trivial, we realize that the edge states of zigzag and armchair nanoribbons cross zero energy at $k_x=0$ and $k_y=0$ coinciding with a gapless and a gapped bulk spectrum, respectively. This difference results on a different localization length of the edge states, becoming fully delocalized, and thus, trivial in the case of zigzag edge states and exponentially localized in the armchair case, as shown in Fig.~\ref{fig:Bulk_CMs}(c).   
In this scenario, the armchair edge states become confined among effectively trivial edges, and the presence or absence of zero energy states become strongly dependent on the length of the armchair edge. Such instability is caused by the presence of the zigzag edge and can be removed using periodic boundary conditions, see further details in App.~\ref{app.stability}. 
Alternatively, we can gap out the bulk nodal point responsible for the observed behavior along the zigzag edges. This will set a finite localization length on the corresponding chiral mode, and thus, reestablish the chiral Majorana mode on the zigzag edges. 

In the rest of the paper, we focus on the effect of quantum confinement along the armchair direction, which allows to open a gap at the nodal points scaling as $1/N_y$, yielding Majorana bound states at the boundaries of the resulting quasi-one dimensional zigzag nanoribbons. 

\begin{figure}
     \centering
     \includegraphics[width=\linewidth]{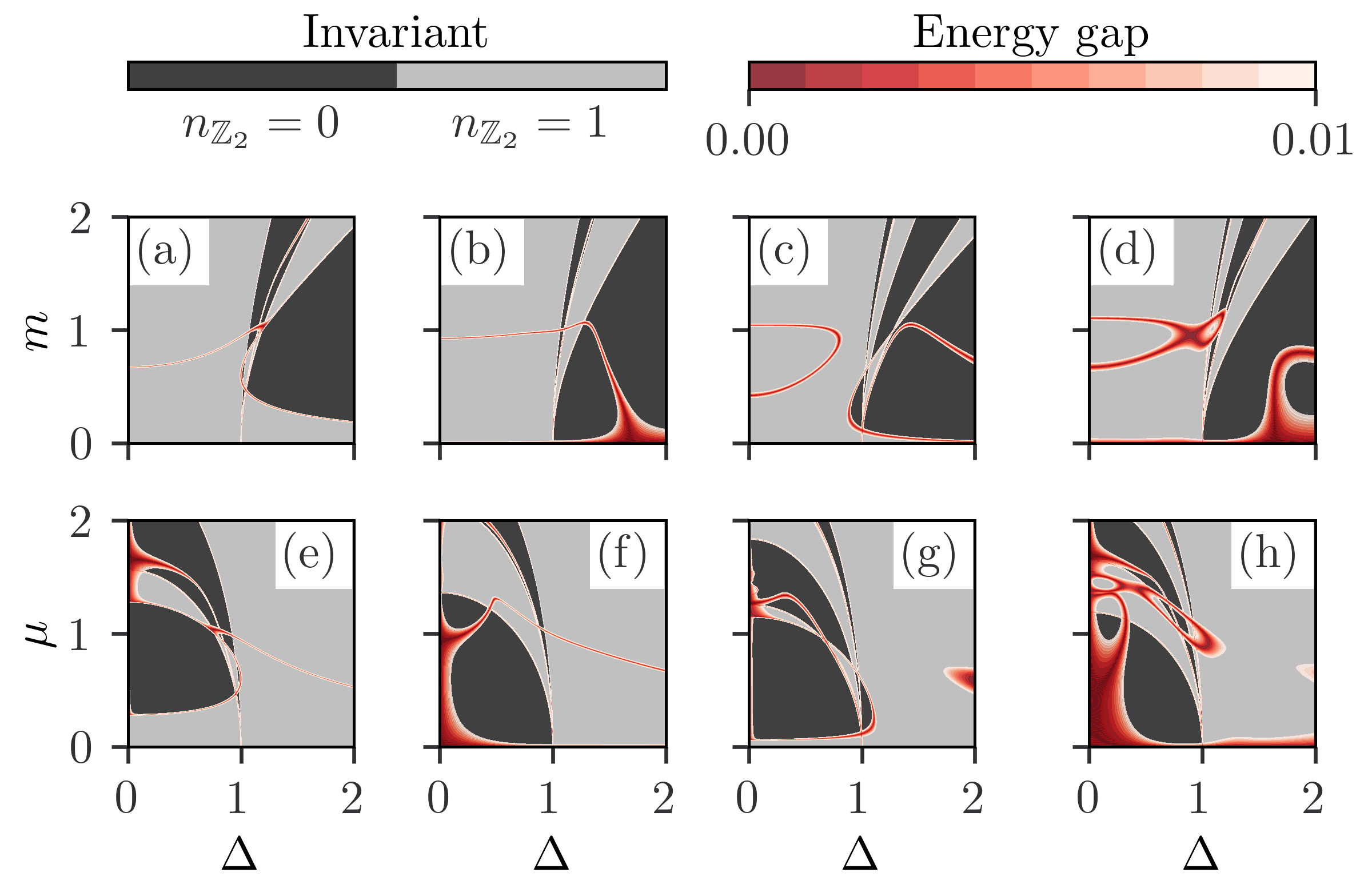}
     \caption{Phase diagram for strips of various widths (respectively $N_y = 4,6,8,10$ going left to right), with $t_2=0.3$; $\mu=0$ in Panels (a)-(d) and $m=0$ in Panels (e)-(h). The light and dark gray areas correspond to regions of the phase diagram where the Pfaffian is equal to 1 (trivial) and -1 (topological) respectively. In the reddish regions the gap of the strip with periodic boundary conditions is close to zero: the darker the shade of red the smaller the gap is. In the resulting dark gray regions MBSs are expected in an open boundary geometry (flake).}
    \label{fig:strip_phase_diag}
\end{figure}

\section{Quasi-one dimensional strip}
\label{Sec:Q1D}
\subsection{Phase diagram}
From now on we consider narrow zigzag nanoribbons. The derivation of the nanoribbon Hamiltonian, together with its Majorana representation is given in App.~\ref{app:nanoribbon}. In order to characterize the topology of the quasi 1D system as a function of the nanoribbon width, one needs to compute the Majorana number, that serves as $\mathbb{Z}_2$ topological index. Alternatively, one could also resort to real-space topological invariants, such as those based on the Majorana polarizations~\cite{Sticlet_2012, Maiellaro_2022b} or those based on multipartite entanglement indicators~\cite{Maiellaro_2022a, Maiellaro_2023}. However, given that the system we consider is translational invariant in the absence of disorder, there is no need to resort to such more sophisticated techniques.

As long as the system is gapped, the Majorana number~\cite{Kitaev_2001} is given in terms of the $B$ matrix associated to its Majorana representation, see App.~\ref{app:nanoribbon} and~\onlinecite{zenodo}
\begin{equation}
    \mathcal{M}(B)= \sgn\left \{ \text{Pf}\, B(0)\right\}\sgn\left\{ \text{Pf}\, B(\pi)\right\},
\end{equation}
where the $\Pf B$ denotes the Pfaffian of the matrix $B$, that is skew-symmetric at the TR invariant points $k=0$ and $k=\pi$.

In Fig.~\ref{fig:strip_phase_diag}(a-d) are reported the $m-\Delta$ phase diagrams for nanoribbons with $4,\ 6,\ 8$ and $10$ sites in the vertical direction and $\mu=0$. To the phase diagram defined by the Majorana number (darker regions have $\mathcal M=-1$) is superimposed a density plot of the nanoribbon energy gap. In correspondence of the darker shades of red the nanoribbon gap is closed. By comparing with the bulk phase diagram in Fig.~\ref{fig:Bulk-PD}(a), it is apparent that the nanoribbon becomes topological across the bulk gap closing defined by the line $m=3\sqrt{\Delta^2-t_1^2}$ (i.e. $\Delta_{\text{c},1}$ for $\Theta = m$). In the region to the right of this line, the phase space is divided in topological ($\mathcal M=-1$) and trivial ($\mathcal M=+1$) regions. 
In the original nodal phase sector ($\Delta_\text{c,1}>|\Delta|> \Delta_\text{c,2}$), we observe alternating regions of trivial and topological regimes. 
This alternating behavior increases with the strip width because the confinement potential imposes a smaller level spacing, $\sim 1/N_y$, and therefore, becomes energetically easier to change the number of channels below the Fermi energy as a function of $m$ and $\Delta$. This phenomenon is known as the even-odd effect in the context of topological two-dimensional superconductors~\cite{Reuther2013a}. Moreover, the original gapped regions ($|\Delta|> \Delta_\text{c,2}$) show a more stable behavior, since the gap is larger. However, bears notable differences for $m<1$ or $\mu<1$ between the two cases $N_y=4M$ and $N_y=4M+2$, $M \in \mathbb Z^*$. This difference can be qualitatively expected as an effective interference pattern between the paths connecting the two edges, as in Ref.~\onlinecite{Traverso_2024}.

In Fig.~\ref{fig:strip_phase_diag}(e-h) are reported analogous plots for the case $m=0$. The phase diagrams are now in the $\mu-\Delta$ plane, and must be compared with the bulk phase diagram in Fig.~\ref{fig:Bulk-PD}(b). In this case, the nanoribbon can host a topological phase across the gap closing line with equation $\mu =3\sqrt{t_1^2-\Delta^2}$ (i.e. $\Delta_{\text{c},1}$ for $\Theta = i\mu$). In the region to its left, the phase space is divided in topological ($\mathcal M=-1$) and trivial ($\mathcal M=+1$) regions in a complicated way. Again, the pattern strongly depends on the strip width and, showing notable differences for the two cases $N_y=4M$ and $N_y=4M+2$, $M \in \mathbb Z^*$.

\begin{figure}  
\centering
\includegraphics[width=\linewidth]{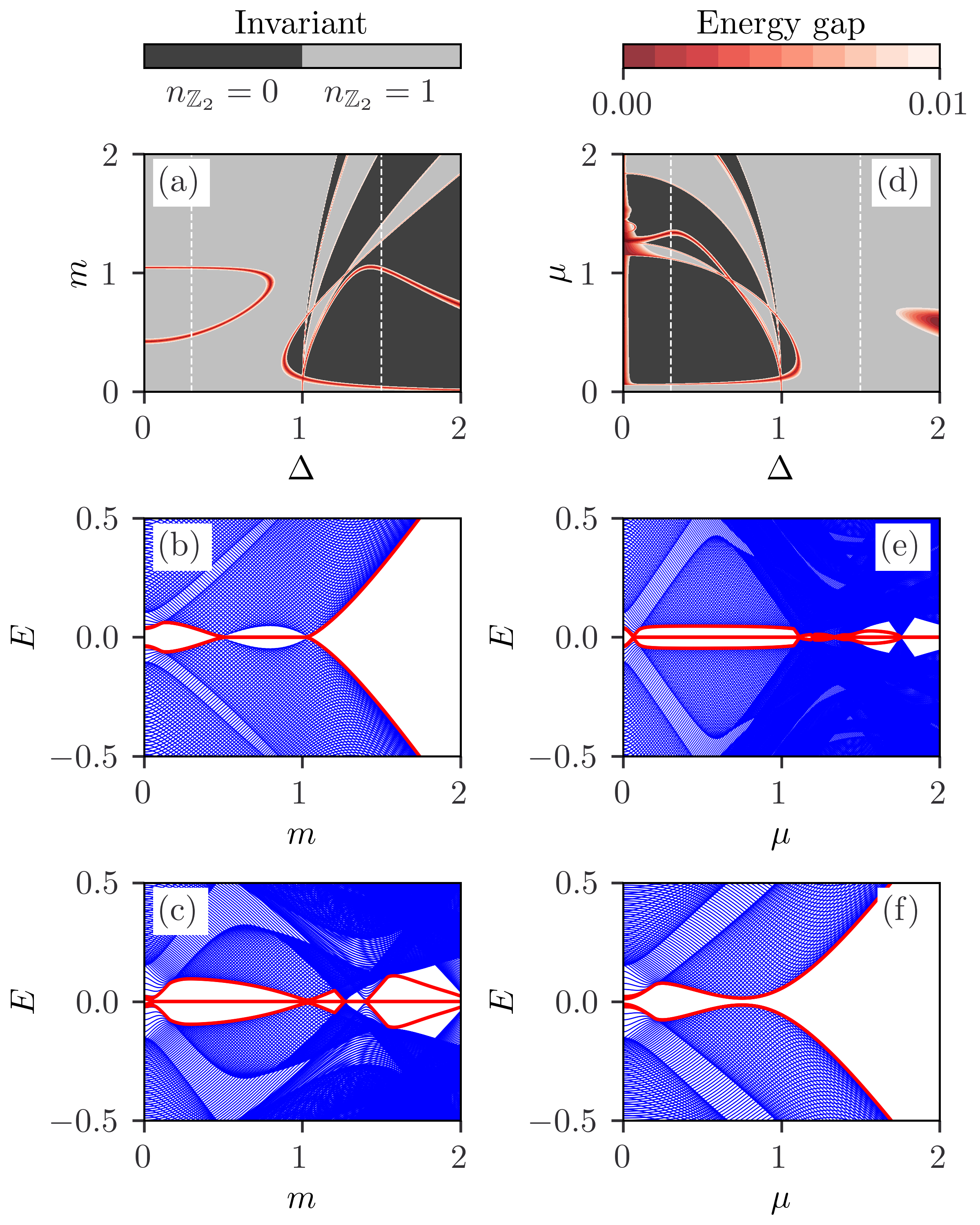}
\caption{Representation of the full PBC spectrum plus the four lowest energy eigenvalues of the OBC spectrum for a nanoribbon of width $N_y=8$ and $N_x=400$ as a function of $m$ ($\mu=0$) in Panel (b,c) and as a function of $\mu$ ($m=0$) in Panel (e,f). In Panels (a,d) are shown zoomed replicas of the phase diagrams of Fig.~\ref{fig:strip_phase_diag}(c) and Fig.~\ref{fig:strip_phase_diag}(g) respectively: the vertical white dashed lines represent the cuts in the parameter space over which the spectra are computed. In particular, the superconducting gap $\Delta$ is set to $0.3$ in Panels (b,e) and to $1.5$ in Panels (c,f); $t_2=0.3$ in all plots.}
\label{fig:Full_spectrum}
\end{figure}

The topological classification yielded by the Majorana number is meaningful only in the regions of the parameter space where the nanoribbons are found to be gapped, that is, away from the darker shades of red in the plots of Fig.~\ref{fig:strip_phase_diag}. In correspondence of the topological regions of the phase diagram, we expect Majorana zero modes to occur at boundaries of a (long) finite flake.

\subsection{Emerging Majorana bound states}
In order to verify the validity of the phase diagrams obtained in the previous subsection, we consider some cuts at fixed values of $\Delta$ and compute the spectra as a function of $m$ and $\mu$ for a strip of given length, both under periodic boundary conditions (PBC) and under open boundary conditions (OBC). Then, we show in Fig.~\ref{fig:Full_spectrum} the full PBC spectrum together with the four lowest energy OBC eigenvalues. 
Here, we can see that the OBC eigenspectrum coincides with the PBC one, except when additional zero energy modes localised at the boundaries are present. Therefore, by plotting just the four lowest energy eigenvalues of the OBC spectrum, we are able to visualize the intervals in $m$ ($\mu$) where only two zero energy in-gap eigenvalues are present. Specifically, intervals in which the four OBC eigenvalues overlap entirely with the PBC spectrum do not host boundary states. In contrast, those where all four OBC eigenvalues are pinned at zero energy, indicate the presence of trivial bound states. Notably, intervals in which the two lowest-energy OBC eigenvalues lie at zero energy, while the next two overlap with the PBC spectrum, indicate the presence of a Majorana bound states localized at the nanoflake ends.

In Fig.~\ref{fig:Full_spectrum} are reported such plots for nanoribbons with $N_y=8$, accompanied by a zoomed replica of the corresponding phase diagram, as a reference. By comparing the plots of the spectra with the corresponding cuts in the phase diagram (indicated by white dashed lines) one finds an exact correspondence. For the case $\mu=0$ the cut at $\Delta/t_1 = 0.3$ [Fig.~\ref{fig:Full_spectrum}(b)] is found to be fully trivial: Indeed, the bound states occurring between the two gap closings are a total of four and as such they are not Majorana bound states. The cut at $\Delta/t_1 = 1.5$ instead [Fig.~\ref{fig:Full_spectrum}(c)], presents pairs of MBSs in correspondence of the gapless topological regions of the phase diagram in Fig.~\ref{fig:Full_spectrum}(a). An analogous correspondence can be found between the spectra of Fig.~\ref{fig:Full_spectrum}(e,f) and the phase diagram of Fig.~\ref{fig:Full_spectrum}(d), obtained for $m=0$. The main difference is that the fully trivial cut here is the one at $\Delta/t_1 =1.5$, as expected.

It is worth noting that for small values of $\Delta$ in the case $\mu=0$, the gap closings and the presence/absence of (four) bound states in an open boundary zigzag flake substantially matches the results reported in Ref.~\onlinecite{Traverso_2024}. Indeed, we have shown is subsec.~\ref{sec:model} that the superconducting term does not directly interfere with the Haldane coupling, since $f(\mathbf k)=0$ at the Dirac points. Thus, for $\mu=0$ and small values of $\Delta$, the BdG spectra results in a trivial doubling of those that one would obtain in absence of superconductivity-- at least with this specific superconducting pairing --and the physics described in~\cite{Traverso_2024} still qualitatively holds.

\subsection{Conductance}
It is well-established that the emergence of a MBS at a junction between a normal metal and a topological superconductor can lead to a sharp quantization of the zero bias conductance to the value $G=\frac{\de I}{\de V}|_{V=0}=2\frac{e^2}{h}$~\cite{Law2009a, Beenakker_2009}. {Although not conclusive}~\cite{Prada2020a}, the zero bias conductance {needs} to be regarded as a {signature} of the presence of Majorana zero modes in our system, to be compared with the phase diagram of Fig.~\ref{fig:strip_phase_diag}.

\begin{figure}
    \centering
    \includegraphics[width=\linewidth]{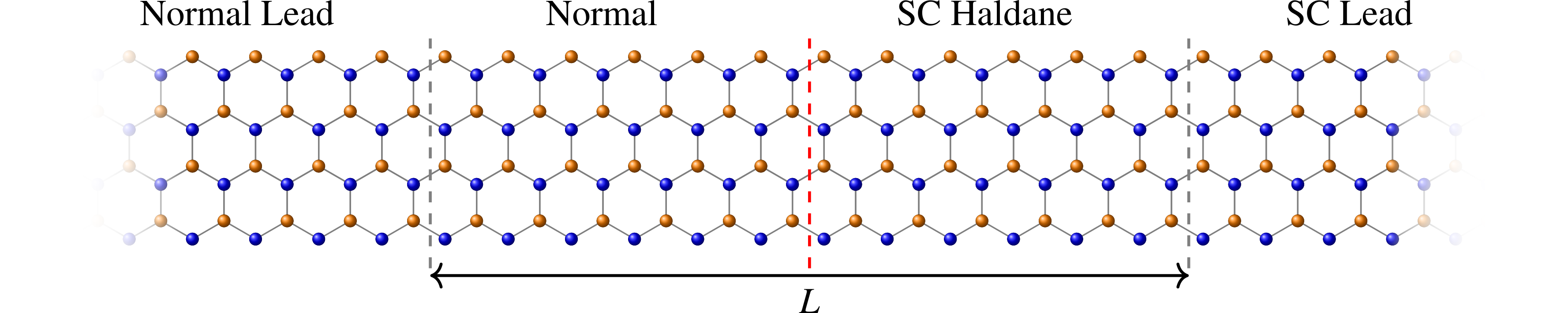}
    \caption{Scheme of the junction between normal part (spinless graphene) and the superconducting Haldane part. The scattering region is normal in the left half and superconducting in the righ part. $L$ denotes the total length of the scattering region.}
    \label{fig:junction_scheme}
\end{figure}

The setup we consider consists of a normal lead (spinless graphene) and a superconducting Haldane lead, connected respectively to the left and to the right of a hybrid scattering region, as illustrated in Fig.~\ref{fig:junction_scheme}. We take the total length of the scattering region as $L=16$ and divide it in two parts of equal length, the left one being normal and the right one hosting the superconducting Haldane coupling.
Moreover, we fix $m=0$ and $\mu/t_1 =0.2$ in the left lead and in the left half of the scattering region; thus the conductance is computed as a function of the parameters of the superconducting side only.
Finally, an on-site random potential is added in the scattering region, of the form
\begin{equation}
    H_{\text{dis.}} = \sum_{i\in \text{S.R.}} w_i c^\dagger_{i}c_{i}, 
\end{equation}
with $w_i$ randomly extracted from the interval $[-V_0/2,+V_0/2]$. It is worth noting that various forms of local perturbations could have been considered (e.g. disorder in the hopping parameters, or lattice defects). Here, however, disorder is introduced primarily to remove all trivial bound states from zero energy. This enables a clearer identification of zero-bias conductance quantization as a signature of a Majorana zero mode (MZM) within our model, and facilitates direct comparison with the phase diagrams shown in Fig.~\ref{fig:strip_phase_diag}.
Moreover, MZMs have been shown in numerous previous theoretical studies to be resilient against all types of short-range disorder~\cite{Prada2020a}, and therefore, we expect the same behavior in our system.

\begin{figure}
    \centering
    \includegraphics[width=\linewidth]{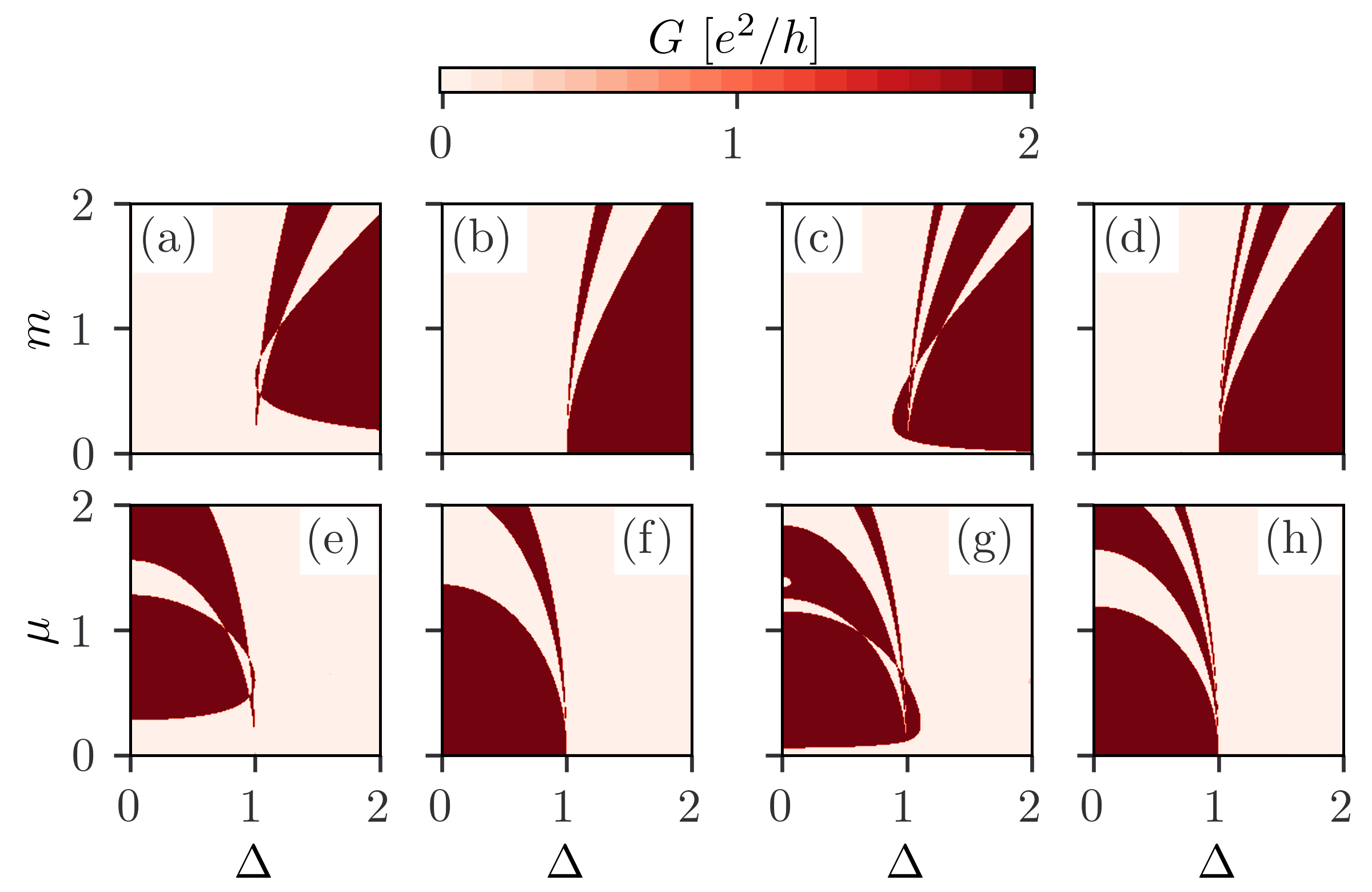}
    \caption{Density plots for strips of various widths (respectively $N_y = 4,6,8,10$ going left to right), representing the disorder-averaged conductance as a function of $\Delta$ and $m$ for $\mu=0$ in Panels (a)-(d), and as a function of $\Delta$ and $\mu$ for $m=0$ in Panels (e)-(h). In all plots the Haldane coupling is set to $t_2=0.3$. The conductance is computed by averaging over 10 different realization of random on-site disorder in the scattering region, with maximum magnitude $V_0=0.05$. The conductance is given in units of $e^2/h$.}
    \label{fig:G-DP_noise}
\end{figure}

To compute the conductance we use the non-equilibrium Green's function method~\cite{Caroli_1971}, which is briefly recalled in section~\ref{sec:NEGFM} and the software used can be found in Ref.~\onlinecite{zenodo}. We work at zero temperature. In Fig.~\ref{fig:G-DP_noise} are shown density plots of the conductance obtained by averaging over 10 different disorder realization, with maximum intensity $V_0/t_1=0.05$. The correspondence with the phase diagrams of Fig.~\ref{fig:strip_phase_diag} is striking: the comparison between the two sets of plots clearly shows that in the presence of a Majorana bound state in the superconducting Haldane zigzag nanoribbon, the conductance remains firmly quantized to $2\frac{e^2}{h}$, while this is not the case in the presence of trivial bound states. We underline once more that in the absence of disorder a quantization of the zero-bias conductance could also be due to the presence of trivial bound states sitting exactly at zero energy. This disorder free case is explicitly shown and discusssed in App.~\ref{app.cond}. Similar conductance phase diagrams arise when we switch the boundary conditions, i.e.~ZZ$\leftrightarrow$AC, see App.~\ref{app.stability}. However, in this occasion the conductance looses its quantization when the width of the strips is larger than $N_x\gtrsim 2\times 30+1$ atoms, resulting from the reduction of the confinement potential.

\section{Conclusions}\label{Sec:conclusions}
In this contribution, we have considered the Haldane model enriched with a {nearest-neighbour equal spin pairing} superconducting term. We have  studied both the {bulk} 2D phase diagram, which encodes interesting physics {\it per se}, and the quasi-1D phase diagram in the context of zigzag nanoribbons.

{The 2D phase diagram of the Haldane model in presence of BCS superconductivity, and in the absence of chemical potential or staggered mass, had been studied before}~\cite{Zhang_2019}.
{However, it is only when either of these is present that a nodal topological superconducting phase emerges. Indeed, the critical lines delimiting the topological region stem from the point $\Delta=t_1,\ m=0$ ($\Delta=t_1,\ \mu=0$), and extend in the $m-\Delta$ ($\mu-\Delta$) plane.}
We have characterized this phase both by computing the proper topological invariant and by analysing the spectrum of wide nanoribbons with armchair and zigzag edges. We have found that chiral Majorana modes, connecting the nodal point projections in the nanoribbon 1-dimensional Brillouin zone, are present in both configurations. However, those hosted by zigzag nanoribbons are effectively trivial due to degeneracy with the bulk bands at the TR-invariant momenta. This effective distinction in the topology of the edge states of the nodal phase has two main consequences: First, in a rectangular-shaped flake, resulting in alternating zigzag-armchair edges, corner modes occur. Second, if one considers thin zigzag nanoribbons, in which the bulk bands are gapped due to quantum confinement, the edge states cease to be fully delocalized in the bulk and are thus re-established. 

In order to explore the topological properties of the model in this quasi-1D setup, we have derived the phase diagram in the $m-\Delta$ ($\mu-\Delta$) plane by computing the Majorana number. In comparison to the bulk phase diagram, here the emerging patterns exhibit a rather intricate structure. This complexity arises from the interplay between the equal-spin superconducting pairing and the staggered mass (chemical potential), in the presence of finite-size effects. In particular, due to the geometry of the underlying honeycomb lattice, confinement effects in this system couple the Majorana edge modes in a totally non-trivial way, resulting in the complicated patterns reported in Fig.~\ref{fig:strip_phase_diag}.
Remarkably, the quasi-1D phase diagrams present some features that are well known in the emerging field of finite size topology of gapped phases~\cite{Lee_2010, Shen_2011, Cook_2023, Traverso_2024}: a reentrant character, with multiple phase transitions; a marked dependence on the nanoribbon width, with the number of phase transitions increasing with the width; a clear even-odd character with respect to the nanoribbon width. However, in the case of this article, such behavior emerges out of a gapless 2D model.

When considering open boundary flakes with parameters belonging to the topological region of the phase diagram, quasi-0D Majorana bound states occur at the two ends of the system. We endorse our results by computing the zero bias conductance at N-S junctions and finding an excellent agreement with the phase diagram yielded by the Majorana number: Indeed, whenever the superconducting side of the junction is in the topological phase, the conductance is found exactly quantized to $2e^2/h$. Moreover, in the presence of disorder, the conductance sharply drops to zero in the trivial phase, while preserving its value in the topological one.

In conclusion, we have investigated the topological phase diagram of the NN-ESP proximitized Haldane model. In two dimensions, we have found a nodal topological superconducting phase, with a chiral Majorana mode propagating along the edges of armchair and zigzag nanoribbons with cylindrical boundary conditions. Interestingly, this topological phase has an unstable character since the localization length of the chiral mode on the zigzag edges is infinite. Thus, when we bring this Hamiltonian onto a rectangular-shape lattice with alternating zigzag and armchair edges, corner states close to zero energy arise. Furthermore, we have found that quantum confinement in one of the directions can stabilize the topological phase by opening gaps on the bulk nodal points, leading to the emergence of Majorana modes in the constricted setup.

Our analysis is mainly a proof of concept that nanostructuring nodal topological superconductors can lead to the emergence of Majorana bound states. Indeed, given a generic (point) nodal topological superconductor, in a nanoribbon geometry the edge modes connect the nodal point projections (\textit{i.e.}~the points of the 1-dimensional Brillouin zone where the bulk gap closes). Moreover, by confining such system in a thin nanoribbon geometry, quantum confinement will gap out the bulk bands. Therefore, whenever the bulk bands are gapped out faster than the edge states-- which is indeed the case for the model we considered --these may still hybridize and give rise to a lower-dimensional topological phase diagram. This will most likely be akin to that presented here and, more generally, present the general features proper of a confined topological phase~\cite{Shen_2011, Cook_2023, Traverso_2024}.

That being said, the model we inspected might still find direct relevance for proximitized honeycomb materials with an intense spin-orbit coupling and in-plane Zeeman fields. Candidate materials meeting these requirements could be represented by bismuthene on SiC~\cite{Reis2017a} and germanene~\cite{Bampoulis_2023}. In fact, the first naturally exhibits topological defects of few nanometers width within the bulk of the material~\cite{Stuehler2022a}, and the second has been recently manifactured with atomic level precision in arrays of narrow zigzag-terminated nanoribbons~\cite{Klaassen_2025}, showing topological features unique to the quasi-1D limit~\cite{Traverso_2024}. Both bismuthene and germanene may hence represent promising platforms for the realization of Majorana fermions, being less affected from disorder with respect to spin-orbit coupled nanowires.

\appendix

\begin{widetext}

\section{Bulk Hamiltonian}
\label{app.SN1}
The Bogoliubov de Gennes Hamiltonian is readily obtained by going in $\mathbf{k}$-space and rewriting as
\begin{equation}
	H = \frac 12 \sum_{\mathbf k} \Psi^\dagger(\mathbf k) \mathcal H_{\text{BdG}}(\mathbf k) \Psi(\mathbf k),
\end{equation}
where the spinor $\Psi(\mathbf k)$ is defined as
\begin{equation}
    \Psi(\mathbf k)=
	\begin{bmatrix}
		a(\mathbf k)\\
		b(\mathbf k) \\
		a^\dagger(-\mathbf k) \\
		b^\dagger(-\mathbf k)
	\end{bmatrix},
\end{equation}
and the operators $a(\mathbf k)$ and $b(\mathbf k)$ are obtained by Fourier transforming the corresponding real space operators
\begin{equation}
	a_{ln}= \dfrac{1}{\sqrt{N_x N_y}}\sum_{\mathbf k}e^{i\mathbf k\cdot \mathbf R_{ln}}a(\mathbf k), \qquad b_{ln}= \dfrac{1}{\sqrt{N_xN_y}}\sum_{\mathbf k}e^{i\mathbf k\cdot \mathbf R_{ln}}b(\mathbf k), \quad \mathbf R_{ln}= l \mathbf a_1+n \mathbf a_2.
\end{equation}
Starting from Eq.~\eqref{eq:real_space_ham} in the main text, one finds
\begin{equation}
	\mathcal H_{\text{BdG}}(\mathbf k) =
	\begin{bmatrix}
		m+\mu+2t_2 S(\mathbf k)  & t_1 f(\mathbf k) &0 &\Delta f(\mathbf k) \\
		t_1 f^{\ast}(\mathbf k)& -m+\mu-2t_2 S(\mathbf k)&-\Delta f(-\mathbf k) & 0 \\
		0& -\Delta f^\ast(-\mathbf k) &-m-\mu+2t_2 S(\mathbf k) & -t_1 f(\mathbf k) \\
		\Delta f^\ast(\mathbf k) & 0 &-t_1 f^{\ast}(\mathbf k) & m-\mu-2t_2 S(\mathbf k)
	\end{bmatrix},
\end{equation}
where $f(\mathbf k)= ( 1 + e^{-i\mathbf k \cdot \mathbf a_1}+e^{-i\mathbf k \cdot \mathbf a_2})$ and $S(\mathbf k) = (-\sin(\mathbf k \cdot \mathbf a_1) + \sin(\mathbf k \cdot( \mathbf a_1-\mathbf a_2))+\sin(\mathbf k \cdot \mathbf a_2))$.

The BdG Hamiltonian can be expressed in terms of tensor product of Pauli matrices as follows
\begin{multline}
	\mathcal H_{\text{BdG}}(\mathbf k) =
	m \tau_z\otimes \sigma_z
	+\mu
	\tau_z \otimes \mathbb I
	+
	2t_2S(\mathbf k)
	\mathbb I\otimes \sigma_z+
	\\
	t_1 \Re(f(\mathbf k))
	\tau_z\otimes\sigma_x
	-t_1 \Im(f(\mathbf k))
	\tau_z\otimes \sigma_y
	-\Delta \Re(f(\mathbf k))
	\tau_y\otimes\sigma_y
	-\Delta \Im(f(\mathbf k))
	\tau_y\otimes \sigma_x,
    \label{eq:BdG_bulk_Ham}
\end{multline}
where the $\tau$ Pauli matrices act on the particle-hole degree of freedom, while the $\sigma$ ones act on the sublattice degree of freedom. If either $\mu=0$ or $m=0$, the Hamiltonian has inversion symmetry $\Gamma$, given by
\begin{equation}
    \Gamma H_{\text{BdG}}(\mathbf k)\Gamma = H_{\text{BdG}}(-\mathbf k).
    \label{eq:inv_sym}
\end{equation}
with $\Gamma = \tau_z\otimes \sigma_x$ for $m=0$, and $\tau_y\otimes\sigma_y$ for $\mu =0$. This implies that, for either $m=0$ or $\mu=0$, $\varepsilon(\mathbf k) = \varepsilon(-\mathbf k)$. Thus, in the presence of both inversion symmetry and particle-hole symmetry ($\mathcal{C}=\mathcal{K} \tau_x \otimes \sigma_0$), we find a symmetric BdG spectrum with respect to zero energy. 


\subsection{Nodal points}
\label{app.nodal}
We now want to derive the bulk phase diagram.
In light of the concluding remark of App.~\ref{app.SN1}, in the following we focus on the case $\mu =0$ alone. We will then we map our conclusions to the $m=0$ case by exchanging $\mu \leftrightarrow m$ and $t_1\leftrightarrow \Delta$. Starting from the analytic expression of the energy bands reported in Eq.~(\ref{eq:bands_mu_0}), one can identify the gapped and gapless regions of the bulk phase diagram. 

To this aim, we need to solve the equation
\begin{equation}
    \left[m^2+4t_2^2S^2(\mathbf k) +(t_1^2+ \Delta^2) |f(\mathbf k)|^2\right]^2=[4mt_2S(\mathbf k)]^2 +[2m\Delta]^2|f(\mathbf k)|^2+[2t_1\Delta|f(\mathbf k)|^2]^2.
\end{equation}

First, one notices that at the Dirac points ($\mathbf{k}_\text{D} = \pm \frac{4\pi}{3}(1,0)$), $f(\mathbf k)=0$. Thus we retrieve the Haldane gap closing condition,
\begin{equation}
	m = \pm 2t_2S(\mathbf{k}_{\text{D}}) = \pm  3\sqrt{3} t_2.
\end{equation}
Thus, the equal spin superconducting pairing considered here, being proportional to $f(\mathbf k)$ in momentum space, does not compete with the Haldane mass at the Dirac point. In other words, a superconducting gap cannot be opened at the Dirac points in the present model.

With some manipulation, we can recast the above equation as
\begin{equation}
    [m^2- 4t_2^2S^2(\mathbf k)]^2+8t_2^2(t_1^2+\Delta^2)S^2(\mathbf k)|f(\mathbf k)|^2+(t_1^2-\Delta^2)^2 |f(\mathbf k)|^4 +2m^2(t_1^2-\Delta^2)|f(\mathbf k)|^2=0.
\end{equation}
In this form, it is clear that away from the Dirac point the gap cannot close for $\Delta < t_1$. Sufficient condition to have gap closing for $\Delta \geq t_1$ is that $S(\mathbf k)=0$. In fact, when this is the case the above equation reduces to
\begin{equation}
    [m^2+(t_1^2-\Delta^2) |f(\mathbf k)|^2]^2=0 \implies m = \pm \sqrt{(\Delta^2-t_1^2)}|f(\mathbf k)|,
\end{equation}
with $\mathbf k$ such that $S(\mathbf k)=0$. One can readily check that $S(\mathbf k)=0$ on the line $(0,k_y)$ and on those related to this one by $C_3$ symmetry. Crucially, on this line
\[
    |f(0,k_y)|^2= 5+4\cos(\sqrt{3}k_y/2),
\]
so that in the $\mathbf{k}$-region over which $S(\mathbf k)$ cancels out, $|f(\mathbf k)|$ is comprised between $1$ and $3$. Thus, for $m>0$, in the region of the parameter space defined by
\begin{equation}
    \sqrt{(\Delta^2-t_1^2)}<m<3\sqrt{(\Delta^2-t_1^2)},
\end{equation}
the system is gapless. Specifically, the gap closes at the six nodal points solving the system
\begin{equation}
    \begin{cases}
        &S(\mathbf k) = 0, \\
        &|f(\mathbf k)|^2 = \dfrac{m^2}{\Delta^2-t_1^2}.
    \end{cases}
\end{equation}

As anticipated, we can draw similar conclusions for the case $m=0$, by simply exchanging $m$ with $\mu$ and $t_1$ with $\Delta$. By doing so, we conclude that the lines
\begin{equation}
	\mu =\pm 3\sqrt{3} t_2,
\end{equation}
are still gapless (Haldane model gap closing condition). Moreover, the gap is also closed in the region of the phase space defined by
\begin{equation}
    \sqrt{(t_1^2-\Delta^2)}<\mu<3\sqrt{(t_1^2-\Delta^2)}.
\end{equation}
Here, the six nodal points where the gap closes are obtained by solving the system
\begin{equation}
    \begin{cases}
        &S(\mathbf k) = 0, \\
        &|f(\mathbf k)|^2 = \dfrac{\mu^2}{t_1^2-\Delta^2}.
    \end{cases}
\end{equation}

We note in passing that if $t_2=0$ then it is not necessary to restrict to the values of $\mathbf k$ for which $S(\mathbf k)=0$ in order to have gap closing away from the Dirac points. Then, given that $|f(\mathbf k)|$ can assume any value between $0$ and $3$ over the whole Brillouin zone, in this case the system becomes gapless for $|m|<3\sqrt{(\Delta^2-t_1^2)}$ if $\mu =0$ and for $|\mu|<3\sqrt{(t_1^2-\Delta^2)}$ if $m=0$.

\subsection{Zigzag nanoribbon}
\label{app:nanoribbon}
In order to describe a zigzag nanoribbon we start from the Hamiltonian in Eq.~\eqref{eq:real_space_ham} of the main text and impose PBC along the $\mathbf{a}_1$ direction only. The Fourier transformation is thus defined as
\begin{equation}
	a_{ln}= \dfrac{1}{\sqrt{N_x}}\sum_{k}e^{ikx_{l}}a_{n}(k), \qquad b_{ln}= \dfrac{1}{\sqrt{N_x}}\sum_{k}e^{ikx_{l}}b_{n}(k),
\end{equation}
where $x_{l}= l$ (since we set $a=1$). The Fourier transformed Hamiltonian is
\begin{equation}
	\begin{split}
		H &= \sum_k \sum_n t_1((1+e^{-ik})a^\dagger_{n}(k) b_{n}(k) +a^\dagger_{n}(k) b_{n-1}(k) +\text{h.c.} ) \\
		&+( \Delta\sum_{k}\sum_{n}(1 + e^{-ik})a^\dagger_{n}(k) b^\dagger_{n}(-k) + a^\dagger_{n}(k) b^\dagger_{n-1}(-k) +\text{h.c.} )\\
		&+t_2\sum_{k}\sum_{n} -2\sin(k) a^\dagger_{n}(k) a_{n}(k) + (i(e^{-ik}-1)a^\dagger_{n}(k) a_{n+1}(k)+ \text{h.c.}) \\
		&+ t_2\sum_{k}\sum_{n} 2\sin(k) b^\dagger_{n}(k) b_{n}(k) + (-i(e^{-ik}-1)b^\dagger_{n}(k) b_{n+1}(k)+ \text{h.c.})\\
		&+\mu\sum_{k}\sum_{n}(a^\dagger_{n}(k) a_{n}(k)+ b^\dagger_{n}(k) b_{n}(k)) \\
		&+m\sum_{k}\sum_{n}(a^\dagger_{n}(k) a_{n}(k)- b^\dagger_{n}(k) b_{n}(k)).
	\end{split}
\end{equation}
For future convenience, we recast the nanoribbon Hamiltonian in Majorana representation. We start by adopting the following notation for the creation and destruction operators
\begin{equation}
    a_n = c_{2n-1}, \quad b_n = c_{2n}.
\end{equation}
With this notation, the Hamiltonian becomes
\begin{equation}
	\begin{split}
		H &= \sum_k \sum_n t_1((1+e^{-ik})c^\dagger_{2n-1}(k) c_{2n}(k) +c^\dagger_{2n-1}(k) c_{2n-2}(k) +\text{h.c.} ) \\
		&+( \Delta\sum_{k}\sum_{n}(1+e^{-ik})c^\dagger_{2n-1}(k) c^\dagger_{2n}(-k) + c^\dagger_{2n-1}(k) c^\dagger_{2n-2}(-k) +\text{h.c.} )\\
		&+t_2\sum_{k}\sum_{n} -2\sin(k) c^\dagger_{2n-1}(k) c_{2n-1}(k) + (i(e^{-ik}-1)c^\dagger_{2n-1}(k) c_{2n+1}(k)+ \text{h.c.}) \\
		&+ t_2\sum_{k}\sum_{n} 2\sin(k) c^\dagger_{2n}(k) c_{2n}(k) + (-i(e^{-ik}-1)c^\dagger_{2n}(k) c_{2n+2}(k)+ \text{h.c.})\\
		&+\mu\sum_{k}\sum_{n}(c^\dagger_{2n-1}(k) c_{2n-1}(k)+ c^\dagger_{2n}(k) c_{2n}(k)) \\
		&+m\sum_{k}\sum_{n}(c^\dagger_{2n-1}(k) c_{2n-1}(k)- c^\dagger_{2n}(k) c_{2n}(k)).
	\end{split}
\end{equation}
We now define the Majorana operators in terms of the $c_j$ operators as
\begin{equation}
	\gamma_{2\alpha-1}(k) = i[c^\dagger_\alpha(-k) - c_\alpha(k)], \qquad \gamma_{2\alpha}(k) = c^\dagger_\alpha(-k) + c_{\alpha}(k).
\end{equation}
By inverting the above definition, it is possible to rewrite the Hamiltonian in the form
\begin{equation}
	H = \frac{i}{4}\sum_k\sum_{\alpha,\beta} B_{\alpha,\beta}(k)\gamma_{\alpha}(-k)\gamma_{\beta}(k),
    \label{eq:maj_rep}
\end{equation}
where the $B$ matrix (non-zero) entries are given by
\begin{align}
	&B_{4\alpha,4\alpha} = B_{4\alpha-1,4\alpha-1} = -B_{4\alpha-2,4\alpha-2}=-B_{4\alpha-3,4\alpha-3} = - 2it_2\sin(k), \\
	&B_{4\alpha,4\alpha+4} = B_{4\alpha-1,4\alpha+3} = -B_{4\alpha-2,4\alpha+2} = - B_{4\alpha-3,4\alpha+1}= -t_2(e^{-ik}-1), \\
	&B_{4\alpha+4,4\alpha} = B_{4\alpha+3,4\alpha-1} = -B_{4\alpha+2,4\alpha-2} = - B_{4\alpha+1,4\alpha-3}= +t_2(e^{+ik}-1), \\
	&B_{4\alpha-2,4\alpha-3} =-B_{4\alpha-3,4\alpha-2}= (\mu+m), \\
	&B_{4\alpha,4\alpha-1} = -B_{4\alpha-1,4\alpha} =(\mu-m), \\
	&B_{4\alpha-2,4\alpha-1} = t_1[1+e^{-ik}]-\Delta [1+e^{-ik}],\\
	&B_{4\alpha-1,4\alpha-2} = -t_1[1+e^{+ik}]+\Delta [1+e^{+ik}],\\
	&B_{4\alpha-3,4\alpha} = -t_1[1+e^{-ik}]-\Delta [1+e^{-ik}],\\
	&B_{4\alpha,4\alpha-3} = t_1[1+e^{+ik}]+\Delta [1+e^{+ik}],\\
	&B_{4\alpha-2,4\alpha-5} = -B_{4\alpha-5,4\alpha-2}= (t_1-\Delta),\\
	&B_{4\alpha-3,4\alpha-4} = - B_{4\alpha-4,4\alpha-3} = -(t_1+\Delta).
\end{align}
In this representation, the (BdG) spectrum can be obtained by diagonalizing the matrix $iB(k)$.

\end{widetext}

\section{Conductance}\label{sec11}

\subsection{Transport formalism}
\label{sec:NEGFM}
Here we briefly recall the Non Equilibrium Green's Function Method for the computation of the current~\cite{Caroli_1971,Cuevas_1996} and specialize it to the case at hand. Consider a left lead (L) and a right lead (R) attached to a central region (C). Assuming the two leads are uncoupled, the Hamiltonian of the full system can be written as
\begin{equation}
    H = 
    \begin{pmatrix}
        H_{LL} & V_{LC} & 0 \\
        V_{CL} & H_{CC} & V_{CR}\\
        0 & V_{RC} & H_{RR}
    \end{pmatrix},
    \label{eq:NEGF_Ham}
\end{equation}
where $H_{LL}$ ($H_{RR}$) is the Hamiltonian of the left (right) lead and $H_{CC}$ the Hamiltonian of the central scattering region. The terms coupling the leads with the scattering region satisfy $V_{LC}=V_{CL}^\dagger,  \ V_{RC}=V_{CR}^\dagger$.
The advanced single-particle matrix Green's function is defined by
\begin{equation}
    [\omega-H]G^a = \mathbb I,
\end{equation}
with $\omega = E-i\eta$, with $\eta=10^{-7}t_1$. Given the Hamiltonian in Eq.~\eqref{eq:NEGF_Ham}, the above equation can be rewritten as
\begin{equation}
    \begin{pmatrix}
        \omega-H_{LL} & -V_{LC} & 0 \\
        -V_{CL} & \omega-H_{CC} & -V_{CR}\\
        0 & -V_{RC} & \omega-H_{RR}
    \end{pmatrix}
    \begin{pmatrix}
        G^a_{LL} & G^a_{LC} & 0 \\
        G^a_{CL} & G^a_{CC} & G^a_{CR}\\
        0 & G^a_{RC} & G^a_{RR}
    \end{pmatrix}
    =\mathbb I.
\end{equation}
By solving for $G_{CC}^a$ we find
\begin{equation}
    G_{CC}^a = (\omega - H_{CC} - \Sigma^a_L-\Sigma^a_R)^{-1},
\end{equation}
where $\Sigma^a_\alpha = V_{C\alpha}g_{\alpha\alpha}^aV_{\alpha C}$ are the self energies of lead $\alpha=L,R$, with $g_{\alpha\alpha}^a$ the advanced Green's functions of the uncoupled leads
\begin{equation}
    (g_{\alpha\alpha}^r)^\dagger= g_{\alpha\alpha}^a = (\omega - H_{\alpha\alpha})^{-1},\ \alpha=L,R.
\end{equation}
The corresponding retarded Green's function is given by $G_{CC}^r= (G_{CC}^a )^\dagger $.

The aim of the NEGF method is determining the current through the contact due to a constant bias $eV=\mu_L-\mu_R$. In the scenario considered in this paper, the left part is normal and the right part is superconducting and all the matrices are written in the BdG basis. In the following we will assume zero chemical potential in the superconducting part, so that $eV=\mu_L$. The expression for the current across the junction is~\cite{Caroli_1971,Cuevas_1996}
\begin{equation}
        I =  \dfrac{e}{h}\int_{-\infty}^{+\infty} \text{d} E\, \text{Tr} [G_{CL}^{+-}(E)V_{LC}-V_{CL}G_{LC}^{+-}(E)],
\end{equation}
where the trace is computed over the electronic subspace only~\cite{Cuevas_1996} and
\begin{align}
    G_{LC}^{+-} &= g_{LL}^{+-}V_{LC}G_{CC}^a + g_{LL}^r V_{LC} G_{CC}^{+-},\\
    G_{CL}^{+-} &= G_{CC}^{+-} V_{CL} g_{LL}^a + G_{CC}^r V_{CL} g_{LL}^{+-},
\end{align}
with
\begin{equation}
    G_{CC}^{+-} = G_{CC}^rV_{CL}g_{LL}^{+-}V_{LC}G_{CC}^a + G^r_{CC}V_{CR}g_{RR}^{+-}V_{RC}G_{CC}^a,
\end{equation}
and
\begin{align}
    g_{LL}^{+-} &= (g_{LL}^a-g_{LL}^r)F(E,eV), \\
    g_{RR}^{+-} &= (g_{RR}^a-g_{RR}^r)F(E,0).
\end{align}
In the above expressions, $F(E,\mu)$ is a diagonal matrix, with the entries belonging to the electron sector presenting the Fermi function $f(E,\mu)=\frac{1}{e^{\beta(E-\mu)}+1}$ and those belonging to the hole sector presenting the (hole) Fermi function $1-f(-E,\mu)=f(E,-\mu)$. In the present context we are solely interested in the conductance, that is given by
\begin{equation}
    G=\dfrac{\text{d} I}{\text{d} V} = \dfrac{e^2}{h}\int_{-\infty}^{+\infty} \text{d}E\, \text{Tr} \left[\dfrac{\partial G_{CL}^{+-}}{\partial (eV)} V_{LC}-V_{CL}\dfrac{\partial G_{LC}^{+-}}{\partial (eV)}\right],
    \label{eq:cond_formula}
\end{equation}
where
\begin{align*}
    \dfrac{\partial G_{LC}^{+-}}{\partial (eV)} &= (g_{LL}^a-g_{LL}^r)\dfrac{\partial F(E,eV)}{\partial (eV)} V_{LC}G_{CC}^a + g_{LL}^r V_{LC} \dfrac{\partial G_{CC}^{+-}}{\partial(eV)}\\
    \dfrac{\partial G_{CL}^{+-}}{\partial (eV)} &= \dfrac{\partial G_{CC}^{+-}}{\partial(eV)} V_{CL} g_{LL}^a + G_{CC}^r V_{CL} (g_{LL}^a-g_{LL}^r)\dfrac{\partial F(E,eV)}{\partial (eV)},
\end{align*}
with
\begin{equation}
    \dfrac{\partial G_{CC}^{+-}}{\partial(eV)} = G_{CC}^rV_{CL}(g_{LL}^a-g_{LL}^r)\dfrac{\partial F(E,eV)}{\partial (eV)}V_{LC}G_{CC}^a.
\end{equation}
In the case of zero temperature assessed in the present work one has a major simplification. Indeed,
\begin{align}
    &F(E,\mu) = \diag(\theta(\mu-E),\theta(-\mu-E),\ldots)  \implies \\ 
    &\dfrac{\partial F(E,\mu)}{\partial \mu} = \diag(\delta(E-\mu),-\delta(E+\mu),\ldots).
\end{align}
Thus, no actual integral needs to be performed in the numerical implementation {for $T=0$}. Regarding the surface Green's functions of the semi-infinite uncoupled leads ($g^{a/r}_{\alpha\alpha}, \ \alpha=L,R$), these are computed with the infinite recursive Green's function method~\cite{Sancho_1985}.

\begin{figure}[h]
    \centering
    \includegraphics[width=\linewidth]{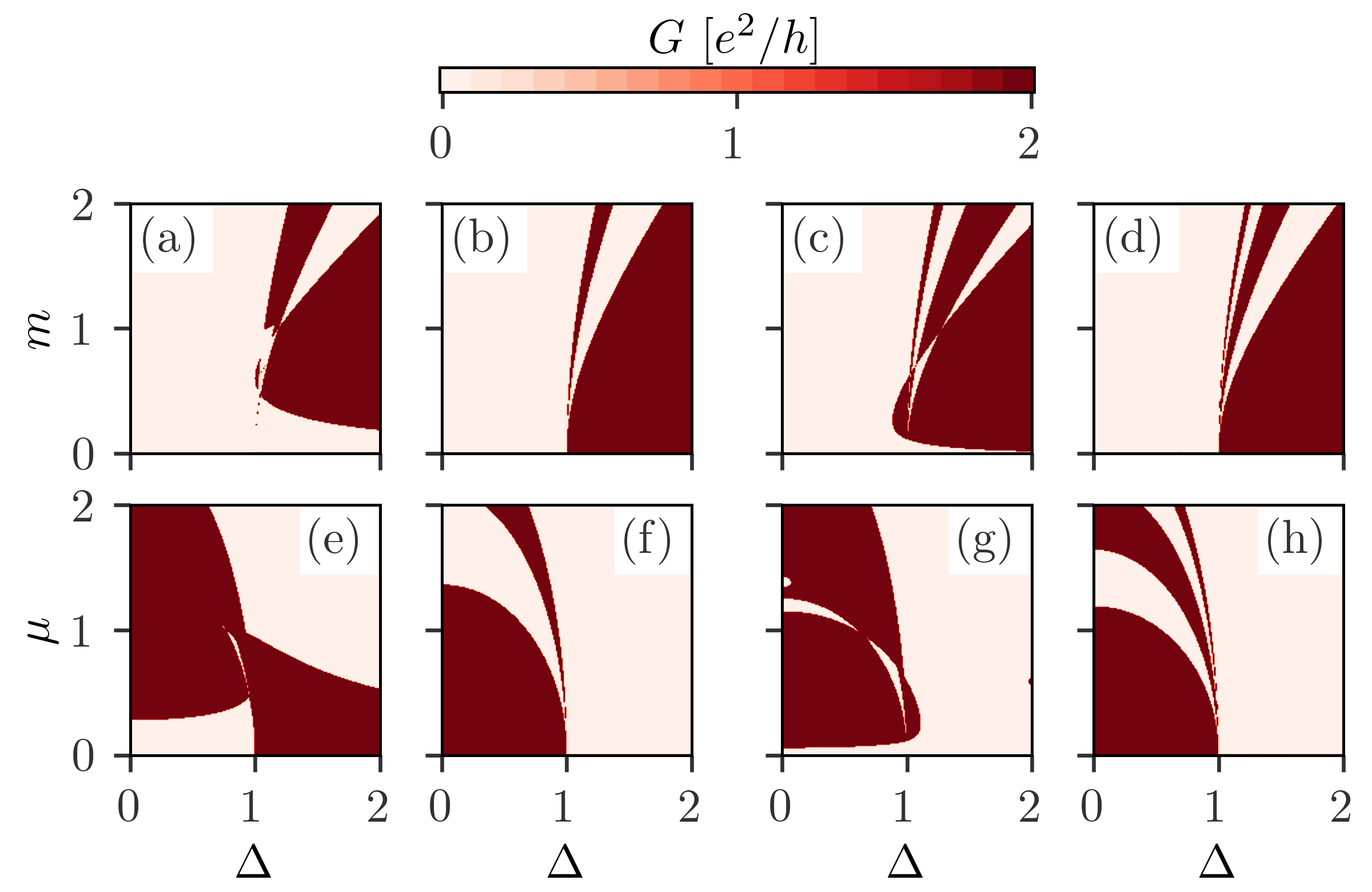}
    \caption{Density plots for strips of various widths (respectively $N_y = 4,6,8,10$ going left to right), representing the conductance as a function of $\Delta$ and $m$ for $\mu=0$ in Panels (a)-(d), and as a function of $\Delta$ and $\mu$ for $m=0$ in Panels (e)-(h). No disorder is added to the scattering region. In all plots the Haldane coupling is set to $t_2=0.3$ and the conductance is given in units of $e^2/h$.}
    \label{fig:G-DP}
\end{figure}

\subsection{Numerical results}
\label{app.cond}

We compute the conductance in the absence of disorder, both for the case $\mu=0$ and for the case $m=0$. The results are shown in Fig.~\ref{fig:G-DP}, where density plots of the conductance as a function of $\Delta$ and $m$ ($\mu$) are reported for different widths. By comparing with the phase diagrams in Fig.~\ref{fig:strip_phase_diag} of the main text, one can see that there is only a partial correspondence: Indeed, though the conductance is exactly $2\frac{e^2}{h}$ in most of the topological regions, it retains such value in some of the trivial ones too. This can be understood by observing that zero energy bound states are present in some of the trivial regions as well.
For example, if we compare Fig.~\ref{fig:G-DP}(g) with Fig.~\ref{fig:strip_phase_diag} of the main text, we see that there is a trivial region where the conductance is quantized to $2\frac{e^2}{h}$.
By comparing with the spectrum along the first cut of Fig.~5(d), shown in Fig.~5(e), we can see that a bound state is present in such trivial region with quantized conductance. However, these are not Majorana zero modes and therefore their energy can be affected by the introduction of random disorder in the scattering region. Thus, as discussed in the main text, we expect that by introducing some form of local disorder in the system the conductance should remain quantized to $2\frac{e^2}{h}$ just in the regions that are topological from the point of view of the superconductivity. This behavior is clearly shown in Fig.~\ref{fig:G-DP_noise} in the main text.

\begin{figure}[tb]
    \centering
    \includegraphics[width=0.85\linewidth]{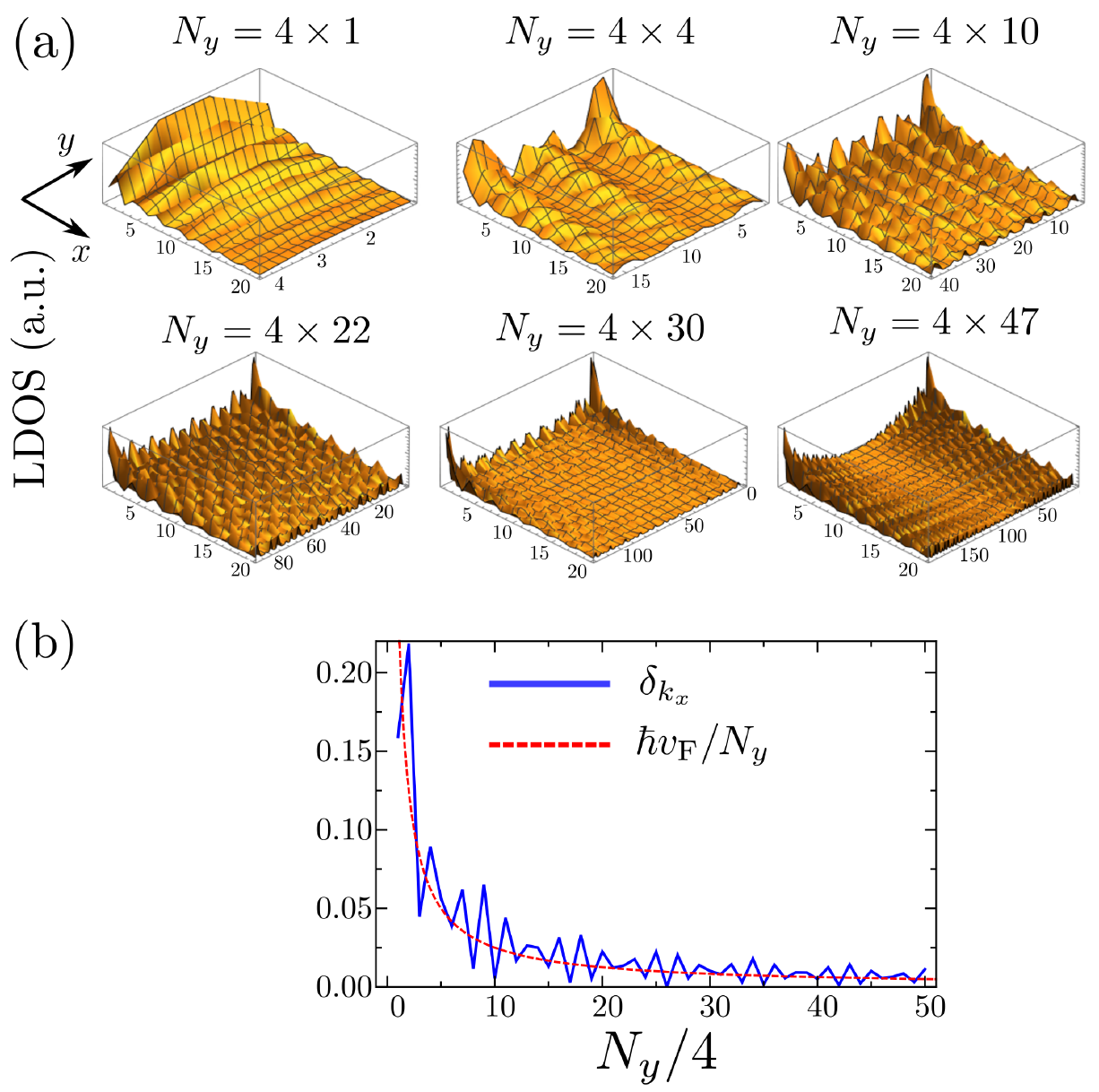}
    \caption{(a) Local density of states for a semi-infinite nanoribbon with armchair (zigzag) boundary conditions along  $y\,$($x$)- direction  for different widths: $N_y=4\times 1\, , 4\times 4\, , 4\times 10\, , 4\times 22\, , 4\times 30\, , 4\times 47$. The parameters used were $\mu=0$, $\Delta=1.4$ and $m=1.75$, which set the Hamiltonian in the nodal phase. (b) Energy of the lowest state in the zigzag nanoribbon $(\delta_{k_x=0})$ as a function of the width. The fitting curve corresponds to the Thouless energy $\hbar v_\text{F}/N_y$.}
    \label{fig:LDOS}
\end{figure}

\section{Wilson loop}
\label{App.Wilson}

The topological invariant $\mathbb{Z}_2$ can be in general computed using the Wilson loop. Here, the idea is to choose a set of Hamiltonians encircling a nodal point in momentum space, with $C\cong S^1$. The resulting set of gapped Hamiltonians in $S^1$ and symmetry class D, exhibit a $\mathbb{Z}_2$ invariant, which can be calculated by means of the line integral over the path $\mathcal{C}$, namely
\begin{align}
    \gamma=-\text{Im}\log \det \left(\prod_j F_j\right) 
\end{align}
where $F_j$ are the overlap matrices
\begin{align}
(F_j)_{mn}=\langle u_m(\varphi_j)|u_n(\varphi_{j+1})\rangle
\end{align}
with $j$ running over discrete values of the path $\mathcal{C}$ in momentum space. The invariant is
\begin{align}
    n\mathbb{Z}_{2}=\frac{\gamma}{\pi}
    \, \text{mod\,2}
\end{align}

\section{Stability analysis of the corner states}
\label{app.stability}

\begin{figure}[tb]
    \centering
    \includegraphics[width=0.95\linewidth]{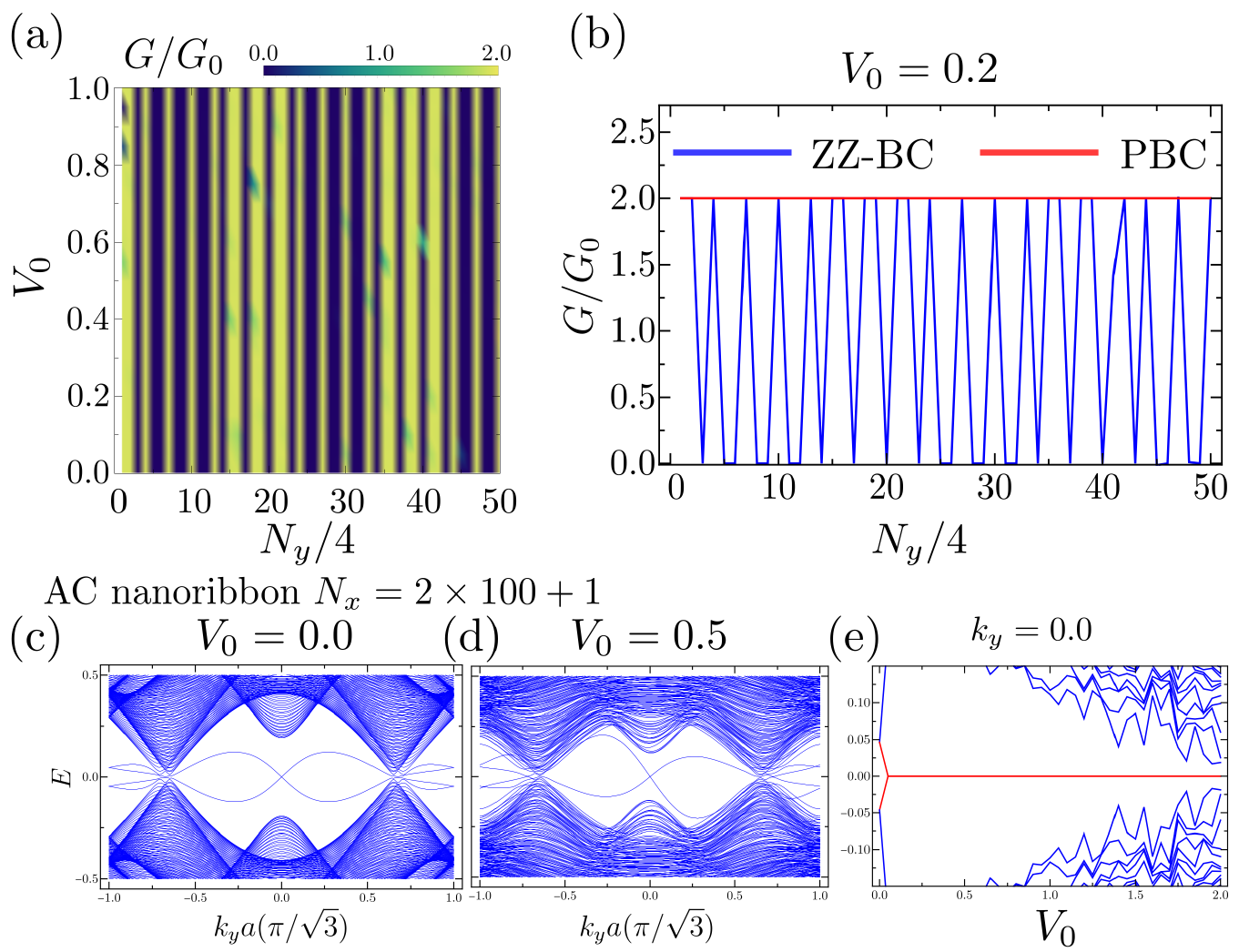}
    \caption{Zero bias conductance as a function of the armchair width and disorder, with $m=1.75$, $\Delta=1.4$ and $t_2=0.3$. (b) Line-cut of panel (a) for a ZZ-boundary condition (ZZ-BC) and periodic boundary conditions (PBC), blue and red curves, respectively. 
    (c,d) Energy vs $k_x$, for an armchair nanoribbon in the nodal phase with disorder strength $V_0=0$ and $V_0=0.5$. (e) Energy as a function of $V_0$ for $k_x=0.0$. We use $\Delta=1.4$, $\mu=0$, $m=1.75$ and $t_2=0.3$ for all panels. In panels (c)-(d) we have used a $N_x=2\times 100+1$  
    }
    \label{fig:ACconductance}
\end{figure}

The so-called corner states arise when the localization length of the zigzag edge states-- set by the bulk gap at the $\Gamma$-point, $\delta_{k_x=0}\propto \hbar v_\text{F}/N_y$, see Fig.~\ref{fig:LDOS}(b)-- is smaller than the width of the sample. In this situation, the edge states along the zigzag edges gap out and the remaining states are confined at the armchair edges and decay exponentially towards the zigzagdirection, see Fig.~\ref{fig:LDOS}(a).\\
From a topological point of view, these edge states are Majorana modes, yielding a quantized zero bias conductance of $G=2e^2/h$ at zero temperature, which is robust against perturbations, see Fig.~\ref{fig:ACconductance}(a,b). From a practical point of view, however, the crossover between both limits is given by the ratio between $\delta_{k_x=0}$ and the temperature.\\
\begin{figure}[htb]
    \centering
    \includegraphics[width=0.75\linewidth]{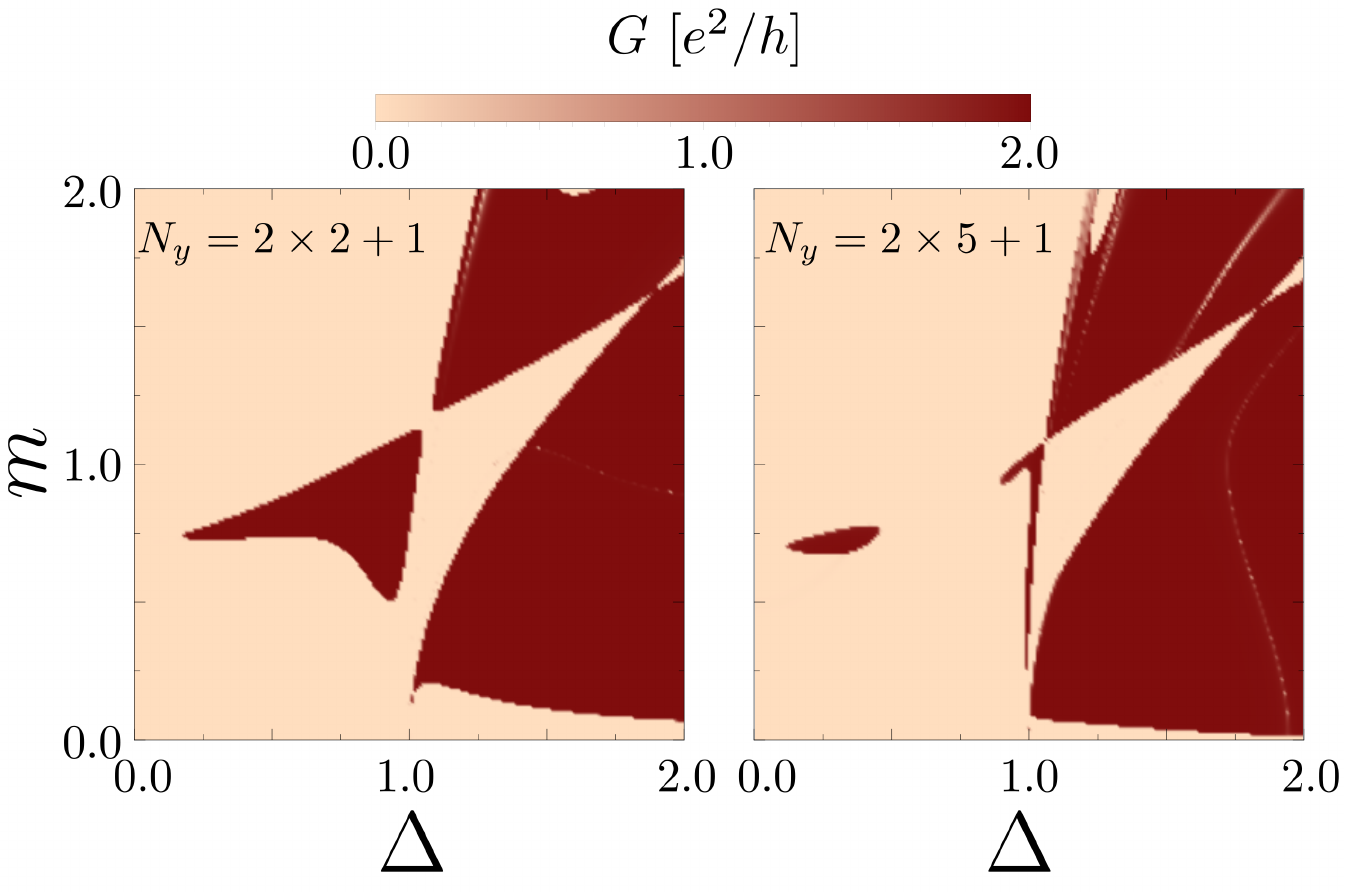}
      \caption{Zero bias conductance as a function of $m$ and $\Delta$ for exchanged boundary conditions, with $t_2=0.3$ and $\mu=0.0$, for widths $N_x=2\times 2+1$ and $N_x=2\times 5+1$. These calculations have been done with disorder strength of $V_0=0.2t_1$.}
    \label{fig:ACconductance_analysis}
\end{figure}
To see this point, we have calculated the zero bias conductance as a function of $N_y$ and disorder strength $V_0$, see Fig.~\ref{fig:ACconductance}(a). Additionally, we have repeated the calculations using periodic boundary conditions to remove the zigzag edge, which is responsible for the instability of the armchair edge states as for certain widths, the energy bulk gap at the $\Gamma$-point is too small for the MBS to be protected. In the first case, we observe that the conductance is quantized for specific widths $N_y$ and is robust against the presence of disorder and zero for other widths, see Fig.~\ref{fig:ACconductance}(a). In contrast, when we impose periodic boundary conditions to remove the zigzag edge, we observe no such $N_y$-dependence, see red curve in Fig.~\ref{fig:ACconductance}(b), showing always a quantized value independently of $N_y$. From these results we can conclude that the corner states resulting from alternating zigzag and armchair edges, have a topological character in a similar way as the MBS observed in the quasi-one dimensional nanoribbons, see panels (c)-(d).\\
To conclude the analysis, we have performed conductance calculations exchanging the boundary conditions relative to the stripe shown in Fig.~\ref{fig:junction_scheme}, that is, with zigzag BC at the NS interface.
In this scenario, we find a similar phase diagrams to those shown in the main text, see Fig.~\ref{fig:ACconductance_analysis}. However, when we increase the width above $N_x\gtrsim 2\times 30+1$ atoms, we observe a lack of conductance quantization.

\begin{acknowledgments}
N. T. Z. acknowledges the funding through the NextGeneration EU Curiosity-driven project ``Understanding even-odd criticality''.
S. T., N. T. Z. and M. S. acknowledge the support from the project PRIN2022 2022-PH852L(PE3) TopoFlags - ``Non reciprocal supercurrent and topological transition in hybrid Nb-InSb nanoflags'' funded by the European community - Next Generation EU within the programme ``PNRR Missione 4 - Componente 2 - Investimento 1.1 Fondo per il Programma Nazionale di Ricerca e Progetti di Rilevante Interesse Nazionale (PRIN)''.
F.D. gratefully acknowledges funding by the Deutsche Forschungsgemeinschaft (DFG, German Research Foundation) under Germany’s Excellence Strategy through the W\"urzburg-Dresden Cluster of Excellence on Complexity and Topology in Quantum Matter ct.qmat (EXC 2147, Project ID 390858490) as well as through the Collaborative Research Center SFB 1170 ToCoTronics (Project ID 258499086)
\end{acknowledgments}

\bibliography{biblio}

\end{document}